\documentclass[reprint,twocolumn,superscriptaddress, amssymb,aps,prx]{revtex4-2}
\usepackage{graphicx}
\usepackage{xcolor} 
\usepackage{amsmath}
\usepackage{hyperref}
\usepackage{tikz}
\usepackage{comment}
\usetikzlibrary{calc,trees,positioning,arrows,chains,shapes.geometric, decorations.pathreplacing,decorations.pathmorphing,shapes,matrix,shapes.symbols}

\newcommand{\ket}[1]{\left| #1\right \rangle}

\begin{document}
\title{Creating and Probing Spin-Squeezed States of Molecules}

\author{Connor M. Holland}
\author{Callum L. Welsh}
\affiliation{Department of Physics, Princeton University, Princeton, New Jersey 08544 USA}
\author{Yukai Lu}
\affiliation{Department of Physics, Princeton University, Princeton, New Jersey 08544 USA}
\affiliation{Department of Electrical and Computer Engineering, Princeton University, Princeton, New Jersey 08544 USA}
\author{David Wellnitz}
\affiliation{Institute for Theoretical Nanoelectronics (PGI-2), Forschungszentrum Jülich, 52428 Jülich, Germany}
\affiliation{Institute for Quantum Information, RWTH Aachen University, 52056 Aachen, Germany}
\affiliation{JILA, NIST and Department of Physics, University of Colorado, Boulder, Colorado 80309, USA}
\affiliation{Center for Theory of Quantum Matter, University of Colorado, Boulder, Colorado 80309, USA}
\author{Xing-Yan Chen}
\affiliation{Department of Physics, Princeton University, Princeton, New Jersey 08544 USA}
\author{Ana Maria Rey}
\affiliation{JILA, NIST and Department of Physics, University of Colorado, Boulder, Colorado 80309, USA}
\affiliation{Center for Theory of Quantum Matter, University of Colorado, Boulder, Colorado 80309, USA}
\author{Lawrence W. Cheuk}
\email{lcheuk@princeton.edu}
\affiliation{Department of Physics, Princeton University, Princeton, New Jersey 08544 USA}

\date{\today}

\begin{abstract}
{Polar molecules are a promising platform for quantum-enhanced sensing and precision tests of fundamental physics, owing to their strong long-range dipolar interactions, broad sensitivity to electromagnetic fields, and sensitivity to potential physics beyond the Standard Model. However, the creation of metrologically useful entangled states in molecular systems has remained elusive. Here, we report the first observation of a class of metrologically useful entangled states --- spin-squeezed states --- in polar CaF molecules trapped in an optical tweezer array. The spin degree of freedom is encoded in rotational levels which are directly  coupled by dipolar exchange interactions. By harnessing appropriate dynamical decoupling schemes we observe up to $3.0(3)\,\text{dB}$ of metrological gain, ($2.2(3)\,\text{dB}$ without measurement correction) from direct exchange interactions. Using Floquet engineering, we further realize richer Hamiltonians that preserve spin squeezing while enabling the development of longer-range quantum correlations. Using site- and spin-resolved measurements we demonstrate that these entangled states enhance sensitivity to both homogeneous and spatially varying fields, and reveal strong non-classical correlations, including bipartite entanglement and Einstein–Podolsky–Rosen steering. Finally, we transfer the spin-squeezed states into long-lived and non-interacting hyperfine states, where the metrological enhancement persists for up to $100\,\mathrm{ms}$. Our results establish molecular optical tweezer arrays as a scalable platform for generating, controlling, characterizing, and storing entangled states of molecules, opening new opportunities for quantum-enhanced sensing and precision tests of fundamental physics.}

\end{abstract}

\maketitle

\section{Introduction}
Spin-squeezed states are entangled states that provide quantum enhancement in sensing~\cite{Kitagawa1993squeezedspin,Pezze2018RMPMetrology}. They host non-classical correlations that reduce quantum projection noise below the Standard Quantum Limit (SQL), the fundamental precision limit for uncorrelated particles. Experimentally, squeezed states of light have provided enhanced sensitivities for gravitational wave detectors~\cite{LIGO2019SqLight}, and spin-squeezed ensembles of atoms have demonstrated quantum-enhanced precision~\cite{Pezze2018RMPMetrology,Leroux2010Squeezing,Hosten2016Squeezing,Bornet2023ScalableSq,Hines2023Squeezing,Lee2025Squeezing,Douglas2025Squeezing}, enabling the realization of quantum-enhanced atomic clocks~\cite{Pedrozo2020ClockSq,Eckner2023Squeezing,Yang2025SqClock}.

Extending quantum-enhanced metrology to polar molecules can open new avenues for both quantum sensing and precision measurement that search for physics beyond the Standard Model~\cite{DeMille2024QSenseReview}. While substantial metrological gains have been achieved in atoms using optical cavities or highly-excited Rydberg states~\cite{Hosten2016Squeezing,Bornet2023ScalableSq}, these experiments are often limited by dissipation~\cite{Leroux2010Squeezing,Hosten2016Squeezing}. Polar molecules naturally realize anisotropic long-range dipolar interactions in long-lived states, enabling squeezing dynamics with little dissipation. In addition, compared to the all-to-all interactions found in atom-cavity systems, the finite spatial range and tensorial structure of dipolar interactions open opportunities to explore fundamentally new regimes of entanglement generation and propagation~\cite{Perlin2020spinsqueeze,Block2024scalablespinsqueeze,kwasigroch2014bose,Comparin2021Robust}, as well as the creation of spatially structured squeezed states with enhanced sensitivity to inhomogeneous fields~\cite{Bilitewski2023BosonicKitaev} and richer many-body dynamics.

Practically, the rich internal structure of molecules enables programmable interactions~\cite{yan2013observation,Li2023tJreverse,Miller2024XYZ,Lu2026SpinChain} and sensing over exceptionally broad frequency ranges. Both interacting and non-interacting sets of molecular states are available~\cite{Ni2018gate,Picard2024entanglement,Holland2025Erasure}, allowing entangled states to be dynamically generated and subsequently stored in states protected from dipolar dynamics. Hyperfine, rotational, and vibrational transitions span frequencies from rf to THz~\cite{Robichaud2026parity,Kondov2019vibcoh}, enabling sensitivity to a wide range of electromagnetic signals. Finally, many molecules are highly sensitive probes of physics beyond the Standard Model. Already, molecules provide many of the best experimental bounds on new symmetry-violating effects and variations of fundamental constants~\cite{Hudson2011YbFEDM,Eckel2013PbO,Cairncross2017EDM,Andreev2018EDM,DeMille2024QSenseReview}. Yet, the same richness that makes molecules so promising also makes them exceptionally challenging to control, and spin-squeezed states in molecular systems have remained unrealized.

Recently, optical tweezer arrays of molecules~\cite{Anderegg2019Tweezer,Holland2023bichromatic,Cairncross2021NaCSRSC} have emerged as a powerful new platform that overcome these challenges and offer full quantum control over individual molecules. These systems allow assembly of low-defect molecular arrays~\cite{Holland2023Entanglement,Holland2025Erasure}, high-fidelity spin-resolved readout~\cite{Anderegg2019Tweezer,Bao2023Entanglement,Holland2023Entanglement,Lu2026SpinChain}, and precise control of coherent dipolar interactions~\cite{Holland2023Entanglement,Bao2023Entanglement,Picard2024entanglement,Ruttley2025Entanglement}, thereby opening the door to creating and microscopically probing spin-squeezed states of molecules.

In this work, we report the first realization of spin-squeezed states in molecules using a 1D tweezer array of polar CaF molecules. We encode spins in two long-lived rotational states that interact via electric dipolar interactions~\cite{yan2013observation,Holland2023Entanglement,Bao2023Entanglement}, which effectively give rise to a $1/r^3$ XX spin model. By dynamically evolving an unentangled product state under XX interactions, we realize spin-squeezed states with up to $3.0(3)\,\text{dB}$ of metrological gain, ($2.2(3)\,\text{dB}$ without measurement correction). 
We then use Floquet Hamiltonian engineering~\cite{yan2013observation,Holland2023Entanglement} to realize a broader class of $1/r^3$ XXZ spin models, enabling programmable control over the generated entanglement dynamics. While the largest metrological gain is obtained at the native XX limit, the engineered XXZ dynamics display metrological gain for all parameters explored while generating longer-range and more spatially structured quantum correlations.

Using single-site spin readout, we directly probe the spatial growth of these correlations and show that the generated states can provide enhanced sensitivity to both uniform and spatially varying fields. Furthermore, we use these correlations to reveal fundamentally quantum properties including bipartite entanglement and Einstein-Podolsky-Rosen (EPR) steering~\cite{Reid2009EPR,Uola2020Steering,Julsgaard2001EPR,Kunkel2018SpatialEPR}, where measurements on one subsystem non-classically influence, or “steer”, the state of another subsystem. Our results represent the first observation of entanglement and EPR-steering in a \textit{many-body} molecular system. Finally, we show how squeezed states, after their creation, can be stored in non-interacting molecular states needed for sensing. Specifically, we transduce XX-squeezed states into a highly coherent and non-interacting hyperfine encoding, preserving quantum enhancement for up to $50\,\text{ms}$ and practical advantage for up to $100\,\text{ms}$.

\section{Overview: Creating and Probing Spin-Squeezed States of Molecules}

\begin{figure}[t]
	{\includegraphics[width=\columnwidth]{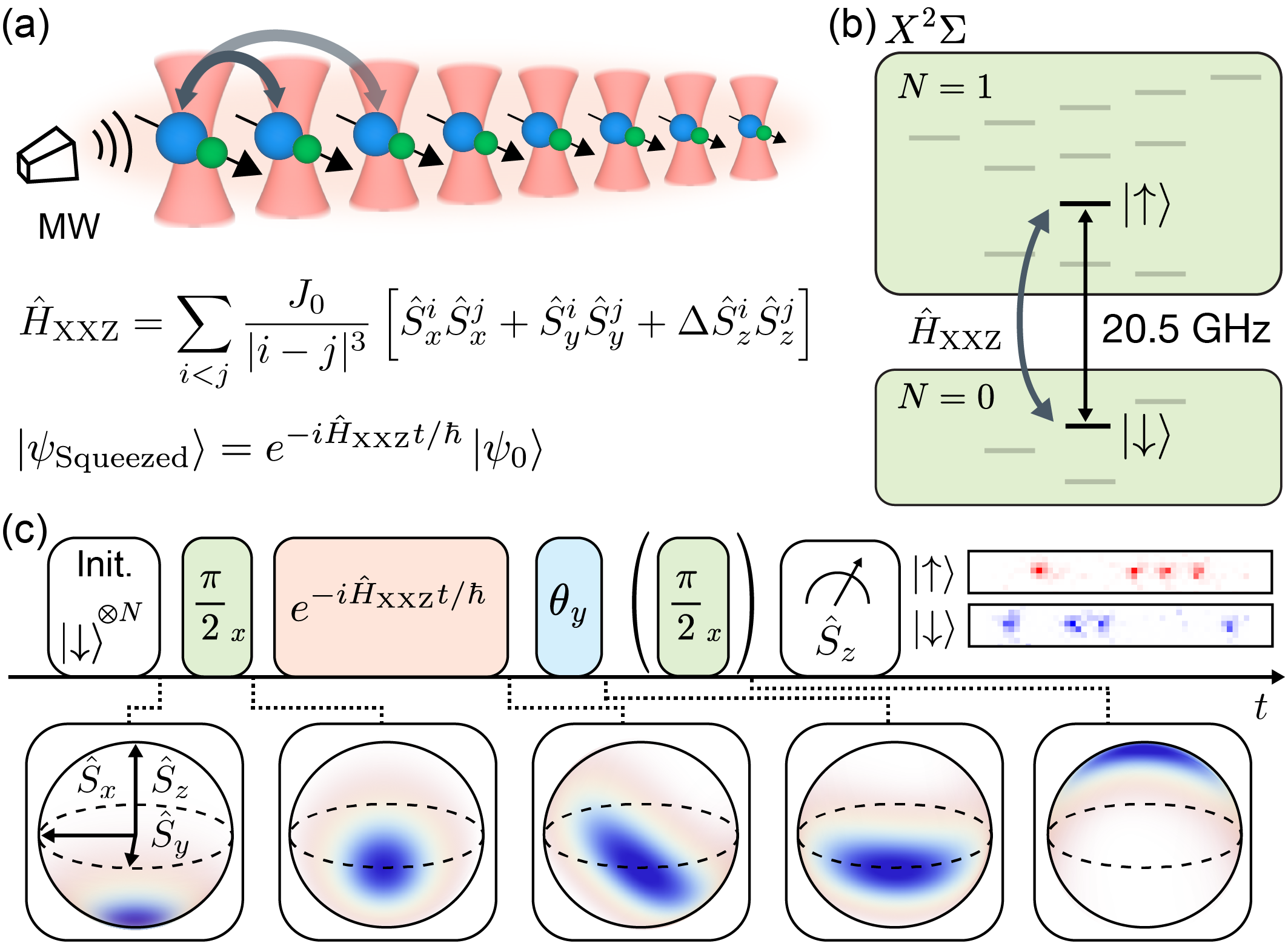}}
	\caption{\label{Fig_1} Spin-Squeezing in a CaF Molecular Tweezer Array. (a) $1/r^3$ XXZ models are realized using Floquet engineering via periodic sequences of global microwave (MW) pulses. (b) CaF rotational qubit encoding. (c) Sequence for dynamically creating spin-squeezed states. Molecules are initialized along the $y$-axis of the collective Bloch sphere via a $\pi/2$ pulse about $x$, after which OAT-like dynamics shear the quantum state and produce squeezing. Rotation pulses applied after squeezing re-orient the state before spin-resolved population detection, which measures $\hat{S}_z$. Specifically, the spin noise $\text{Var}[\hat{S}_\theta]$ is measured after a rotation of angle $\theta$ about the $y$-axis, while the spin length requires an additional $\pi/2$ pulse about the $x$-axis. The bottom panels shows the quantum state on the collective Bloch sphere at various points in the sequence. Shown on the left are exemplary single-shot spin-resolved images. }
	\vspace{-0.2in}
\end{figure}

Our work starts with CaF molecules trapped in a 1D optical tweezer array, where a spin-1/2 degree of freedom is encoded in the states $\ket{\downarrow}=|{X^2\Sigma(v=0,N=0,F=1, m_F=0)}\rangle$ and $\ket{\uparrow}=|{X^2\Sigma(v=0,N=1,F=0, m_F=0)}\rangle$, with $v$, $N$, $F$, $m_F$ denoting vibrational, rotational, and hyperfine quantum numbers, respectively. The array has 8 sites with a uniform spacing of $2.1\,\mu\text{m}$, with each occupied by up to one molecule prepared in the internal state $\ket{\downarrow}$ (Fig.~\ref{Fig_1}(a)). Through rearrangement and measurement-aided state preparation~\cite{Holland2025Erasure}, these arrays are largely free of internal state defects, with an overall preparation fidelity of $f=0.934(5)$, mainly limited by vacancies. The native electric dipolar interactions between molecules give rise to $1/r^3$ spin-exchange interactions (XX model)~\cite{yan2013observation,Holland2023Entanglement,Bao2023Entanglement}, which together with Floquet engineering implemented by periodic microwave pulses addressing the $|{\uparrow}\rangle - |{\downarrow}\rangle$ transition~\cite{Miller2024XYZ,Lu2026SpinChain}, realize a tunable $1/r^3$ XXZ Hamiltonian
\begin{equation*}
	\hat{H}_\text{XXZ}=\sum_{i<j}\frac{J}{|i-j|^3}(\hat{S}_x^i\hat{S}_x^j+\hat{S}_y^i\hat{S}_y^j+\Delta \hat{S}_z^i\hat{S}_z^j),
\end{equation*} 
where $\hat{S}_\beta^i$ are the spin-1/2 operators at site $i$. Because our Floquet sequences distribute the native interactions into exchange terms ($\hat{S}_x^i\hat{S}_x^j$+$\hat{S}_y^i\hat{S}_y^j$) and an Ising term ($\hat{S}_z^i\hat{S}_z^j$) with strength $\Delta$, at fixed molecular separation, $(1+\Delta/2)J$ is a constant~\cite{Miller2024XYZ,Lu2026SpinChain}. For our experiment, $(1+\Delta/2)J=h\times33\,\text{Hz}$ and $\Delta$ is tunable from 0 to $\approx 1.8$.

As first pointed out in Refs.~\cite{Rey2008Squeeze,Perlin2020spin,Block2024scalablespinsqueeze,kwasigroch2014bose}, and measured in Refs.~\cite{Bornet2023ScalableSq,FrankeQuantum2023,Douglas2025Squeezing}, finite-range XXZ models, similar to the well-studied collective Ising model~\cite{Kitagawa1993squeezedspin,Rey2008Squeeze}, are capable of spin-squeezing through one-axis-twisting (OAT) dynamics. An unentangled coherent-spin-state (CSS) initialized on the equator of the Bloch sphere shears, reducing spin uncertainty along a particular spin axis (Fig.~\ref{Fig_1}(c)). This directly reduces quantum projection noise for a Ramsey measurement, but also reduces the collective spin length, which determines the Ramsey fringe contrast, and hence the signal strength. The overall metrological performance is determined by the noise-to-signal ratio encoded by the Wineland squeezing parameter 
$\xi^2_W= \mathcal{N}\,\text{Var}[\hat{S}_{{\bf n}_\theta}]/{\langle\hat{S}_y\rangle ^2}$, where $\hat{S}_{{\bf n}_\theta}=\hat{S}_z\cos(\theta)+\hat{S}_x\sin(\theta)$, $\hat{S}_\beta=\sum_i \hat{S}_\beta^i$ are the collective spin operators, and $\mathcal{N}$ is the particle number. Here, $\text{Var}[\hat{S}_{{\bf n}_\theta}]=\langle \hat{S}_{{\bf n}_\theta}^2\rangle -\langle \hat{S}_{{\bf n}_\theta}\rangle^2$ is the variance for orientation angle $\theta$. For an unentangled ensemble, $\xi^2_W\geq\xi^2_{SQL}$ where $\xi_{SQL}^2=1$ is known as the Standard Quantum limit (SQL). $\xi^2_W<1$ directly indicates quantum metrological gain.

In our experiments, we perform full site and spin-resolved readout, which enable the states of each tweezer (molecule in $|{\uparrow}\rangle$, molecule in $|{\downarrow}\rangle$, and empty site $|{\emptyset}\rangle$) to be measured for every experimental run (Fig.~\ref{Fig_1}(b)). Importantly, this enables post-selection for vacancy-free arrays (see~\cite{Supplement} for details). 
Furthermore, via microwave pulses on the $\ket{\uparrow}$-$\ket{\downarrow}$ transition, the effective readout basis can be oriented along any desired direction on the Bloch sphere. These capabilities provide access to spatial spin correlations, spin variance $\text{Var}[\hat{S}_{\mathbf{n}}]$, and spin length $\langle{|\hat{S}_{\mathbf{n}}|}\rangle$ along any direction $\mathbf{n}$.

\section{Spin-Squeezing with XX Interactions}
We first investigate spin squeezing with XX interactions ($\Delta=0$). We initialize all molecules in $\ket{\downarrow}$ and apply a global $\pi/2$ pulse about the $x$-axis to align all molecules  along the -$y$-axis  $|{\downarrow}\rangle_y = \frac{1}{\sqrt{2}}(|{\uparrow}\rangle -i |{\downarrow}\rangle)$. We suppress noise during the subsequent evolution with an XY8 dynamical decoupling sequence~\cite{Supplement}, such that the molecules evolve according to $\hat{H}_\text{XXZ}$ with $\Delta=0$ for an interaction time $t$.

\begin{figure}[t]
	{\includegraphics[width=\columnwidth]{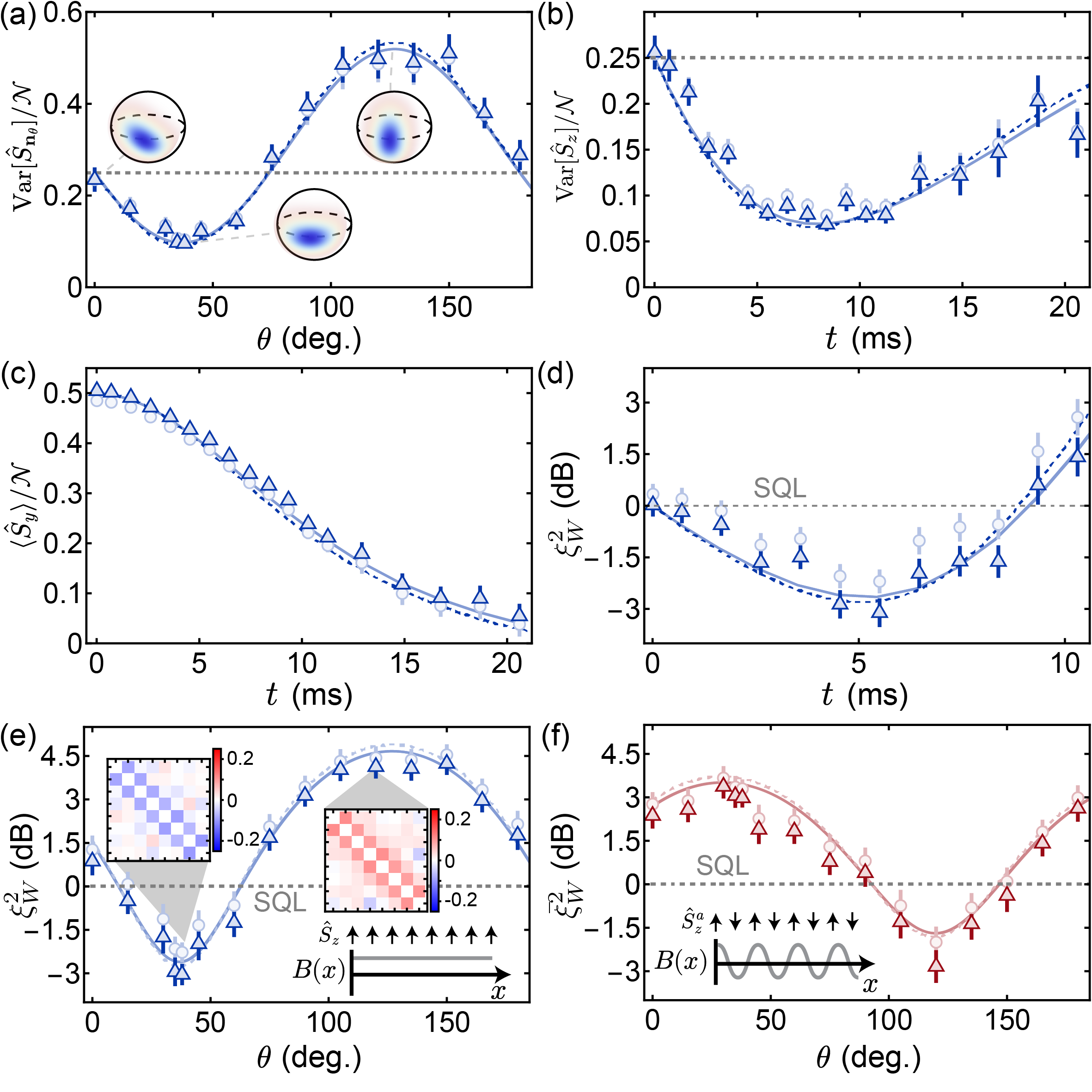}}
\caption{\label{Fig_2} XX Squeezing Dynamics and Metrological Performance. (a) Normalized variance of $\hat{S}_{\bf{n}_\theta}$ versus orientation angle $\theta$, measured after $4.55\,\text{ms}$ of XX squeezing. (b) Normalized variance of $\hat{S}_z$ versus interaction time $t$ at $\theta^*$. (c) Normalized spin length $|\langle \hat{S}_y\rangle |$ versus interaction time $t$. (d) Wineland parameter $\xi^2_W$ versus squeezing duration $t$ at $\theta^*$. (e) $\xi^2_W$ versus readout angle $\theta$. Insets show the spin-spin correlators $\langle{S_z^iS_z^j}\rangle_c$ at $\theta=38^\circ$ (left) and $\theta=120^\circ$ (right), revealing predominantly short-ranged correlations. (f) The staggered Wineland parameter $\bar{\xi}^2_W$ versus readout angle $\theta$. Inset: illustration of a spatially staggered magnetic field with alternating sign on even and odd sites that can be sensed by $\hat{S}^a_z$. For all plots, exact diagonalization theory and our full noise model are shown by the dashed and colored bands, respectively. For all plots, light circles show the raw data without measurement correction, the triangles show data that are measurement-corrected for identification and process errors~\cite{Supplement}.}
	\vspace{-0.2in}
\end{figure}

As a preliminary investigation, we measure the variance of $\hat{S}_{{\bf n}_\theta}$, $\text{Var}[\hat{S}_{{\bf n}_\theta}]$, versus the readout angle $\theta$ near the theoretically optimal squeezing duration of $t=4.55\,\text{ms}$. We find that $\text{Var}[\hat{S}_{{\bf n}_\theta}]$ oscillates versus $\theta$, revealing the asymmetric spin noise profile expected from a spin-squeezed state (Fig.~\ref{Fig_2}(a)). We find good agreement between the measured variances with both a disorder-free exact diagonalization simulation of $\hat{H}_\text{XXZ}$ and simulations including dominant sources of noise~~\cite{Supplement}.

We next investigate both the spin noise $\text{Var}[\hat{S}_z]$ and the collective spin length $|\langle{\hat{S}_y}\rangle|$ versus interaction time $t$. For all subsequent measurements, after evolution with $\hat{H}_\text{XXZ}$, we apply a rotation of $\theta^*$ to align the minimum noise quadrature along the measurement basis ($z$), with $\theta^*$ determined from theory. Fig.~\ref{Fig_2}(b,c) shows $\text{Var}[\hat{S}_z]$ and $|\langle{\hat{S}_y}\rangle|$ versus $t$. We find excellent agreement between the data and an ideal $1/r^3$ XX model, indicating unitary dynamics within experimental uncertainty. To demonstrate quantum metrological gain through squeezing, we extract the Wineland parameter $\xi^2_W$ versus $t$ (Fig.~\ref{Fig_2}(d)). $\xi^2_W$ achieves a minimum of $-3.0(3)\,\text{dB}$ ($-2.2(3)\,\text{dB}$ without measurement correction of identification and process errors), indicating metrological gain. We note that $\xi^2_W<1$ directly shows that the state is also entangled~\cite{sorensenmany2001,Hyllus2012QFIEnt,Toth2012QFIEnt}.

Next, we go beyond the collective spin observables and examine microscopic spatial correlations of the squeezed states, which are directly accessible in our tweezer platform. Unlike collective all-to-all spin models for which spatial correlations are uniform, finite-range squeezed states generically have spatial structure~\cite{Eckner2023Squeezing,Douglas2025Squeezing}. In particular, we examine the connected spin-spin correlators $\langle{\hat{S}_z^i \hat{S}_z^j}\rangle_c = \langle{\hat{S}_z^i \hat{S}_z^j}\rangle-\langle{\hat{S}_z^i}\rangle\langle{\hat{S}_z^j}\rangle$ along the measurement basis. We note that these directly probe \textit{squeezing} correlations, because the variance of a collective spin observable is a sum of all two-spin spatial correlators: $\text{Var}[\hat{S}_z]=\sum_{i\neq j} \langle \hat{S}_z^i \hat{S}_z^j \rangle_c+\sum_i \text{Var}[\hat{S}_z^i]$. Specifically, reduced quantum noise arises from negative $\langle{\hat{S}_z^i \hat{S}_z^j}\rangle_c$ correlations.  

As shown in the insets of Fig.~\ref{Fig_2}(e), $\langle{\hat{S}_z^i \hat{S}_z^j}\rangle_c$ is dominated primarily by nearest-neighbor correlations for the maximally squeezed quadrature ($\theta^*=38^\circ$), in stark contrast with uniform correlations expected in collective squeezed states. We also examine correlations of the anti-squeezed quadrature ($\theta=120^\circ$), where the collective spin noise is maximal. For a pure state, the maximum of $\text{Var}[\hat{S}_z]$ also directly measures the quantum Fisher information, which in turn provides a fundamental limit on the state's sensitivity to global fields~\cite{Pezze2009entanglementqfi,Hyllus2012QFIEnt,Toth2012QFIEnt}. 

We find that $\langle{\hat{S}_z^i \hat{S}_z^j}\rangle_c$ correlations in the anti-squeezed quadrature are positive and also predominantly nearest-neighbor (Fig.~\ref{Fig_2}(f)). By inverting $\hat{S}_z$ at every other site, these positive correlations can be converted to negative correlations, and hence suppressed noise~\cite{CooperGraph2024}. We therefore expect the staggered collective spin operator $\hat{S}_\beta^a = \sum_i (-1)^i \hat{S}_\beta^i$ to be squeezed. Physically, $\hat{S}_\beta^a$ senses spatially oscillating fields with a two-site period. In Fig.~\ref{Fig_2}(f), we plot the equivalent Wineland parameter $\bar{\xi}^2_{W} = \mathcal{N}\,\text{Var}[\hat{S}_{{\bf n}_{\theta}}^a]/\langle\hat{S}_y\rangle^2$ versus readout angle $\theta$. Notably, near the maximal anti-squeezing direction, we find $\bar{\xi}^2_{W}=-2.8(5)\,\text{dB}$ for measurement-corrected data ( $-2.0(5)\,\text{dB}$ without measurement correction), indicating metrological gain. Therefore, the XX-squeezed state displays metrological gain for both uniform and staggered fields, but along two orthogonal directions. Combined with dynamic tweezer rearrangement, this state could allow quantum-enhanced sensing of spatially oscillating fields with arbitrary wavelengths.


\begin{figure}[t]
	{\includegraphics[width=\columnwidth]{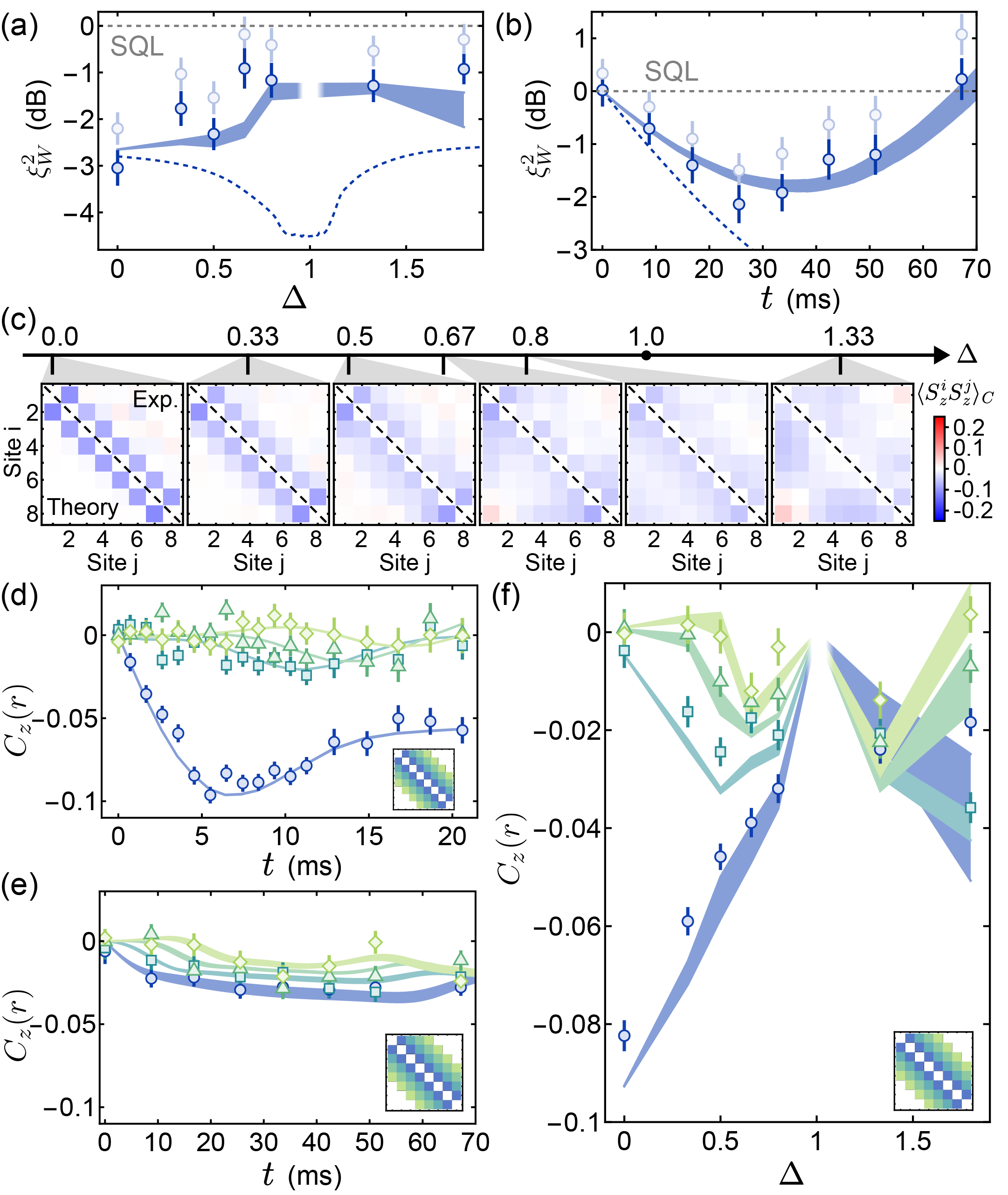}}
	\caption{\label{Fig_3} Metrological Gain and Correlations for XXZ Squeezing. (a) Wineland parameter $\xi^2_W(t^*,\theta^*)$ at the ideal squeezing optimum versus XXZ anisotropy $\Delta$. (b) $\xi^2_W(t,\theta^*)$ versus interaction time $t$ for $\Delta=0.8$. For (a,b), the dashed line shows unitary dynamics from an ideal $1/r^3$ XXZ model. The shaded band shows the predictions of our noise model. (c) Connected spin-spin correlators $\langle{S_z^iS_z^j}\rangle_c$ measured at $t^*$ for different $\Delta$  (upper-right: experiment; lower-left: theory). The diagonals are not shown. (d) The mean spin-spin correlator $C_z(r)$ for $\Delta=0$ versus $t$, where $r$ is the spin separation. (e) $C_z(r)$ versus $t$ for $\Delta=0.8$. (f) $C_z(r)$ versus $\Delta$, revealing extended correlations near the Heisenberg point. For (d,e,f), the insets show the color coding for the separation $r$. For all plots, colored bands indicate our full noise model simulation, and are produced by varying the Floquet micowave pulse detuning by $\pm 500\,\text{Hz}$}. 
	\vspace{-0.2in}
\end{figure}

\section{$1/r^3$ XXZ-Squeezing}
We next investigate spin squeezing under general XXZ interactions with $\Delta\neq0$ implemented through Floquet engineering~\cite{Supplement}. Previous work predicted that close to the Heisenberg point, $\Delta=1$, the almost isotropic interactions keep spins aligned over longer times and distances~\cite{ReyMany2008,Perlin2020spin}. The spin locking improves the attainable spin squeezing, but comes at the cost of longer squeezing durations.

We measure $\xi^2_W(t^*,\theta^*)$ at six values of $\Delta = \{0.33, 0.5, 0.67, 0.8, 1.33, 1.8\}$, tuning through the Heisenberg point. Here, $t^*$ and $\theta^*$ are the optimal squeezing time and rotation angle of $\hat H_{\rm XXZ}$~\cite{Supplement}. As shown in Fig.~\ref{Fig_3}(a), we find that $\xi^2_W<1$ for all values of $\Delta$. However, our results deviate significantly from the ideal $1/r^3$ XXZ model (dashed line), and we observe no improvement in metrological gain over the XX model.

We attribute this difference to noise that accumulates over the long squeezing times. This can be seen in Fig.~\ref{Fig_3}(b), where we plot a complete time trace for $\xi^2_W(t)$ at $\Delta=0.8$. While the data lie close to the dashed line at short times, at longer times the points deviate significantly, and the squeezing reaches its optimal value much earlier. We have developed a physical noise model based on microscopic simulation of the pulsed sequence that incorporates single-particle dephasing arising from, for example, tweezer-intensity fluctuations and finite temperature, finite-temperature-induced local $\hat S^i_z$ inhomogeneities and interaction strength variations, and microwave detuning and phase noise. The latter become increasingly important with the larger numbers of microwave pulses used during the Floquet sequences used to modify the spin couplings~\cite{Supplement}. The numerical simulations are shown as shaded bands in Fig.~\ref{Fig_3}, reflecting uncertainties in the model parameters. The good agreement with experiment indicates that the model captures the dominant noise mechanisms and underscores the high level of coherent control required for Floquet Hamiltonian engineering.

Beyond metrological gain, we probe the spatial structure of the generated entangled states through connected spin correlations $\langle \hat S_z^i \hat S_z^j \rangle_c$ measured at the experimentally identified optimal squeezing time $t^*$~\cite{Supplement} as a function of $\Delta$. As the interactions approach the Heisenberg point ($\Delta=1$), the correlations become increasingly extended across the array (Fig.~\ref{Fig_3}(c)). To quantify this behavior, we extract the distance-resolved correlations
$
C_z(r,t)=\frac{1}{N-r}\sum_i \langle \hat S_z^i \hat S_z^{i+r}\rangle_c(t)
$.
For the native XX dynamics, the squeezing is dominated by short-range nearest-neighbor anti-correlations. In contrast, under Floquet-engineered XXZ dynamics ($\Delta=0.67$), correlations at different distances evolve nearly synchronously and remain closely locked together throughout the dynamics (Fig.~\ref{Fig_3}(d,e)). At the optimal squeezing time, the convergence of $C_z(r,t^*)$ for different $r$ near $\Delta=1$ (Fig.~\ref{Fig_3}(f)) demonstrates the onset of nearly uniform correlations across the system, consistent with emergent all-to-all squeezing dynamics in the proximity of the Heisenberg point. Excellent agreement with the theory model is observed for all the explored parameters.  

\section{Bipartite Entanglement and EPR steering in XXZ-Squeezed States}

The observed spin-squeezing directly implies non-classical correlations~\cite{ma2011quantum}. To go beyond a collective squeezing parameter and resolve how correlations are distributed between two spatially interleaved molecular ensembles, we use site-resolved measurements to determine if correlations between the odd ($A$) and even ($B$) sublattices can suppress joint fluctuations below what is possible for each ensemble alone. 

We directly quantify such bipartite entanglement between odd ($A$) and even ($B$) sites in our system constructing the separability witness $\mathcal{W}=4\text{Var}[\hat{S}_z]\text{Var}[\hat{S}_x^a]/|\hat{S}_y|^2$ from measured spin correlations. $\mathcal{W}<1$ indicates bipartite entanglement~\cite{Fadel2018BECEPR}. As shown in Fig.~\ref{Fig_4}(a), $\mathcal{W}<1$ for all $\Delta$, certifying entanglement between the even/odd partition to $>6\sigma$ of significance for measurement-corrected data ($>3.8\sigma$ without measurement correction).

We note that the corresponding spin observables, $\hat{S}_z$ and $\hat{S}_x^a$
behave as effective continuous-variable quadratures, closely analogous to the two-mode squeezed states that exhibit Einstein--Podolsky--Rosen (EPR) correlations through reduced fluctuations of collective sum and difference observables~\cite{Duan2000Inseparability,Reid2009EPR}. 

We next test whether these correlations are strong enough to demonstrate Einstein--Podolsky--Rosen (EPR) steering. Steering formalizes the original idea by EPR that measurements on one subsystem can predict non-commuting observables of another subsystem more precisely than allowed by any local description of that subsystem alone~\cite{Cavalcanti2009EPRcriterion,Uola2020Steering}. 
We construct an EPR steering witness from spin correlations 
\begin{equation*}
	\mathcal{S}_{A\rightarrow B}=\frac{
		\mathrm{Var}\!\left[ \left( \sum_{i} g_z^i \hat{S}_z^i \right) + \hat{S}_z^B \right]
	\,
		\mathrm{Var}\!\left[ \left( \sum_{i} g_x^i \hat{S}_x^i \right) + \hat{S}_x^B \right]
	}{
		\left| \left\langle \hat{S}_y^B \right\rangle \right|^2 / 4
	}
\end{equation*}
\noindent
with $i\in A$, and $g_i$ left as free-parameters to minimize $\mathcal{S}_{A\rightarrow B}$~\cite{Fadel2018BECEPR}. If $\mathcal{S}_{A\rightarrow B}<1$, measurements in $A$ can steer $B$, i.e.~by measuring either $\sum_{i\in A} g_z^i \hat{S}_z^i$ or $\sum_{i \in A} g_x^i \hat{S}_x^i$, we can predict $\hat S_z^B$ or $\hat S_x^B$ more accurately than any disentangled state could be determined in $B$ alone. In Fig.~\ref{Fig_4}(b), we show both $\mathcal{S}_{A\rightarrow B}$ and $\mathcal{S}_{B\rightarrow A}$ versus $\Delta$ for the maximally squeezed states. For $\Delta=0$ and $\Delta=0.5$, both directions are steerable to at least 2$\sigma$ of significance for measurement-corrected data. Although XXZ squeezing is predicted to produce states capable of EPR steering for all $\Delta$, the delicate EPR correlations are more sensitive to experimental imperfections than bipartite entanglement. Our results here are the first observations of many-body states of molecules displaying bipartite entanglement and EPR steering. 

\section{Transduction of Spin-Squeezing into Non-interacting States}

\begin{figure}[t]
	{\includegraphics[width=\columnwidth]{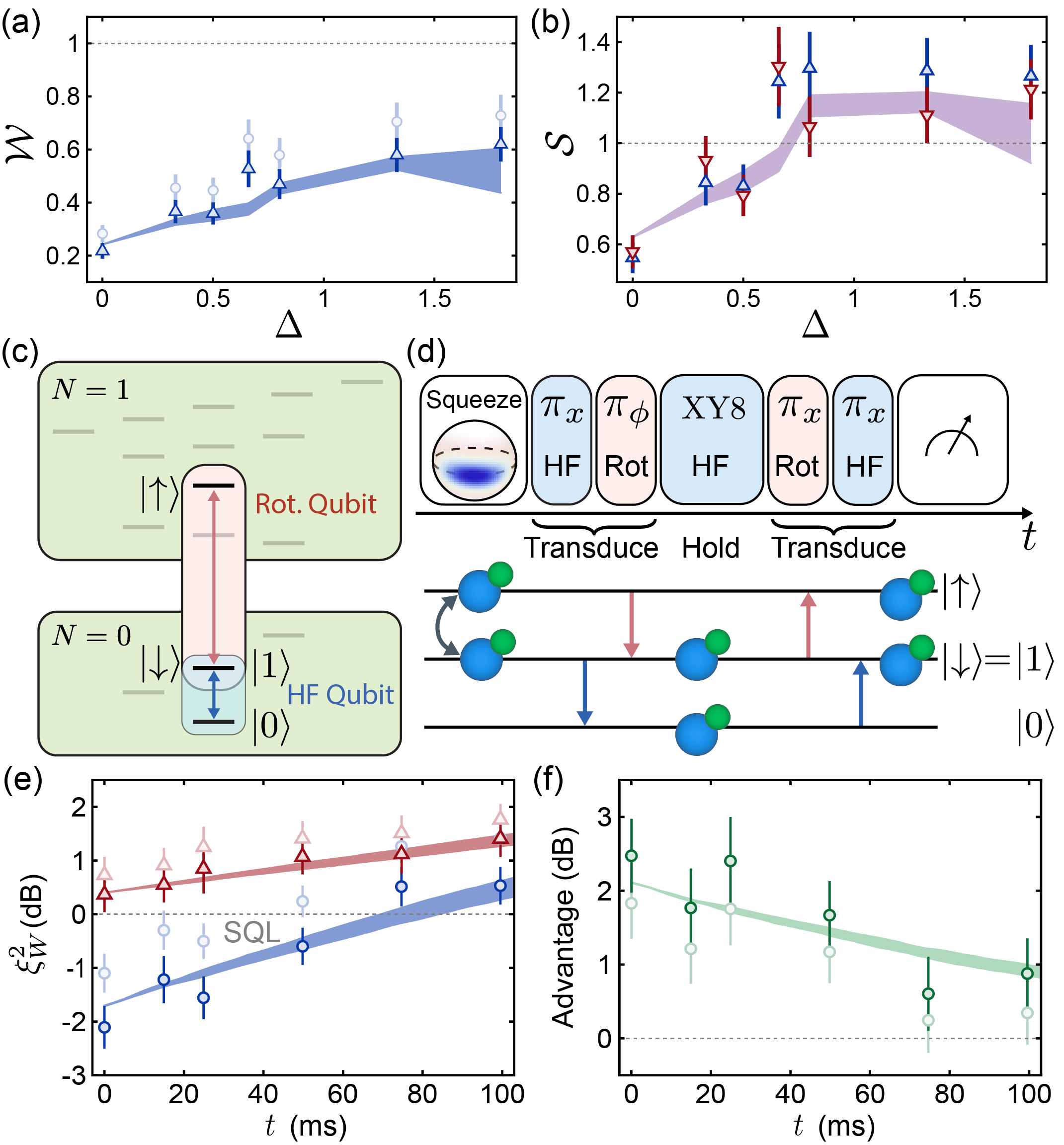}}
	\caption{\label{Fig_4} Bipartite Entanglement, EPR Steering, and Storage of Spin-Squeezed States. (a) Separability witness $\mathcal{W}$ versus $\Delta$. (b) EPR steering witness $\mathcal{S}$ versus $\Delta$ (red: A$\rightarrow$B steering; blue: B$\rightarrow$A steering). (c) Hyperfine qubit encoding $|0\rangle$-$|1\rangle$. (d) Squeezing and transduction protocol. Molecules in the rotational encoding ($|{\uparrow}\rangle$-$|{\downarrow}\rangle$) are squeezed, transferred into the non-interacting hyperfine encoding~\cite{Holland2025Erasure}, and stored. For readout, populations are transferred back to the rotational encoding for spin-resolved detection. (e) Wineland parameter $\xi^2_W$ versus storage time $t$. Red (blue) shows data for an unentangled coherent spin state (spin-squeezed state). (f) Practical metrological advantage of the squeezed state over the coherent spin state ($\xi^2_{W,\mathrm{CSS}}/\xi^2_{W,\mathrm{SSS}}$) as a function of storage time $t$. For all plots, colored bands indicate our full noise model simulation. Theory error bands for hyperfine storage account for filling fraction and decoherence time.}
	\vspace{-0.2in}
\end{figure}
Lastly, we show how our spin-squeezed states, once created, can be stored in non-interacting molecular states. Specifically, we transduce the XX-squeezed state into two hyperfine states from the $N=0$ manifold, $|{0}\rangle=|X^2\Sigma(N=0, F=0, m_F=0)\rangle$ and $|{1}\rangle=|X^2\Sigma(N=0, F=1, m_F=0)\rangle$~\cite{Holland2025Erasure}. This hyperfine encoding has a long measured lifetime and coherence time of $T_1=T_2=730(80)\,\text{ms}$ (with dynamical decoupling)~\cite{Holland2025Erasure}, and selection rules prevent dipolar exchange interactions.

We first create an XX-squeezed state with its squeezed quadrature aligned along $z$. Subsequently, two $\pi$-pulses transduce this state into the hyperfine encoding. After a variable storage time $t$, we measure the Wineland parameter $\xi^2_W$ through coherent global rotations within the hyperfine manifold followed by hyperfine state-selective detection~\cite{Supplement}. For our theoretical model, we add microwave amplitude noise and uniform depolarizing noise account for $T_1 = T_2$. We note that while enhanced sensing requires the state to be squeezed along the equator, the decay of metrological gain is expected to be independent of the state’s orientation in our system since $T_1 = T_2$.

In Fig.~\ref{Fig_4}(e), we show $\xi^2_W$ versus storage time $t$ for both the squeezed state and an unentangled coherent spin state. At $t=0$, we observe a modest degradation in $\xi^2_W$ for both states, which we attribute to technical noise introduced during the transduction process. As a function of $t$, $\xi^2_W$ increases for both states due to decoherence. We find that the stored squeezed state maintains sub-SQL performance ($\xi^2_W < 1$) for over $50\,\text{ms}$. Furthermore, the squeezed state maintains a practical advantage over a coherent spin state for $\approx 100\,\text{ms}$ (Fig.~\ref{Fig_4}(f)). This defines the bandwidth over which the non-classical state provides enhanced metrological performance. 

\section{Summary and Outlook}
In summary, we have realized spin-squeezed states of polar molecules for the first time and demonstrated metrologically useful entanglement in a molecular tweezer array. By engineering tunable dipolar XXZ Hamiltonians, we probed the spatial structure of the resulting correlations, revealing bipartite entanglement and EPR steering between spatially separated molecular ensembles. Our results establish polar molecules as a powerful platform for generating, controlling, and storing distributed entangled states with programmable interactions and long coherence times.

The theoretical models developed here quantitatively capture the dominant experimental imperfections and identify clear pathways toward substantially larger metrological gain in 2D arrays, including optimized XXZ regimes that could outperform the native XX dynamics~\cite{Supplement}. Extending these techniques to larger and higher-dimensional arrays, together with programmable XYZ interactions~\cite{Miller2024XYZ,Lu2026SpinChain}, could enable scalable squeezing, two-axis counter-twisting dynamics~\cite{Koyluoglu2025TACT}, and entangled states tailored for spatially structured sensing protocols~\cite{Periwal2025DipolMagicProp}. Combined with site-resolved control, dynamically reconfigurable geometries, and access to both interacting and protected non-interacting molecular states, these capabilities further open opportunities for differential and multi-parameter sensing schemes based on spatially structured entanglement.

More generally, the tensorial nature of dipolar interactions in molecules provides access to richer many-body dynamics beyond the XXZ models explored here. Additional interaction channels can naturally generate spin-orbit-coupled Hamiltonians and topological bosonic phases~\cite{Syzranov2014SpinOrbital,Bilitewski2023BosonicKitaev}, opening the possibility of combining metrologically useful entanglement with intrinsic robustness against disorder and decoherence.

Incorporating all these quantum-enhanced techniques into precision measurement experiments that exploit the broad spectral bandwidth of molecular internal states and their intrinsic sensitivity to symmetry-violating physics could unlock previously inaccessible gains in sensitivity and bandwidth for the exploration of new physics beyond the Standard Model~\cite{DeMille2024QSenseReview}.

\section{Acknowledgements}
We thank Waseem Bakr, Sarang Gopalakrishnan, and Norm Yao for fruitful discussions. This work was supported by the National Science Foundation under Grant No. 2207518,  Grant No. 2513401, the JILA Physics Frontier Center grant PHY-2317149 and QLCI OMA-2016244; by the Air Force Office of Scientific Research under Grant No. FA9550-24-1-0140, Grant No. FA9550-25-1-0092, and  MURI FA9550-21-1-0069, by the Office of Naval Research under Grant No. N000142512375, Army Research Office Grant  W911NF-24-1-0128, and from the US Department of Energy, Office of Science, Quantum System Accelerator, and the National Institute of Standards and Technology. D. W. acknowledges support by the German Federal Ministry of Research and Technology (BMFTR) project MUNIQC-ATOMS (Grant No. 13N16070) and support from the German Research Foundation (DFG) under Germany’s Excellence Strategy ‘Cluster of
Excellence Matter and Light for Quantum Computing (ML4Q)
EXC 2004/1’ 390534769. C. L. W. acknowledges support from a Princeton Quantum Initiative Graduate Student Fellowship. X.-Y. C. acknowledges support from a Dicke Postdoctoral Fellowship. L. W. C. acknowledges support from the Alfred P. Sloan Foundation under Grant No. FG-2022-19104.

\end{document}


\title{Supplementary Material for\\ ``Creating and Probing Spin Squeezed States of Molecules"}

\author{Connor M. Holland}
\author{Callum L. Welsh}
\affiliation{Department of Physics, Princeton University, Princeton, New Jersey 08544 USA}
\author{Yukai Lu}
\affiliation{Department of Physics, Princeton University, Princeton, New Jersey 08544 USA}
\affiliation{Department of Electrical and Computer Engineering, Princeton University, Princeton, New Jersey 08544 USA}

\author{David Wellnitz}
\affiliation{Institute for Theoretical Nanoelectronics (PGI-2), Forschungszentrum Jülich, 52428 Jülich, Germany}
\affiliation{Institute for Quantum Information, RWTH Aachen University, 52056 Aachen, Germany}
\affiliation{JILA, NIST and Department of Physics, University of Colorado, Boulder, Colorado 80309, USA}

\author{Xing-Yan Chen}
\affiliation{Department of Physics, Princeton University, Princeton, New Jersey 08544 USA}

\author{Ana Maria Rey}
\affiliation{JILA, NIST and Department of Physics, University of Colorado, Boulder, Colorado 80309, USA}

\author{Lawrence W. Cheuk}
\email{lcheuk@princeton.edu}
\affiliation{Department of Physics, Princeton University, Princeton, New Jersey 08544 USA}

\date{\today}

\maketitle

\tableofcontents
  
\section{Experimental Methods}
The experimental apparatus used to prepare CaF molecules in optical tweezers has previously been described in Refs.~\cite{Lu2021Ring,Holland2023bichromatic,Holland2023Entanglement,Li2024blueMOTCaF,Holland2025Erasure,Lu2026SpinChain}. In this work, each experimental run begins with a pulsed CaF molecular beam created using a cryogenic buffer gas source~\cite{Hutzler2012CBGB}. The molecules are then laser-slowed~\cite{Truppe2017chirp} and trapped in a DC magneto-optical trap~\cite{Truppe2017Mot,Li2024blueMOTCaF} within an ultrahigh vacuum chamber. Subsequently, with the aid of sub-Doppler laser-cooling ($\Lambda$-cooling)~\cite{Cheuk2018Lambda,Anderegg2019Tweezer,Holland2023bichromatic,Holland2023Entanglement}, they are transferred into a far-detuned optical dipole trap (ODT) formed with far-detuned $1064\,nm$ light, and then transported to the focal plane of the optical tweezers. The optical tweezer array is formed by passing laser beams of $780\,\text{nm}$ light through a microscope objective, which creates tightly focused spots with a Gaussian waist of $\approx 730\,\text{nm}$. The tweezer locations are controlled by an acousto-optical deflector (AOD) driven by an arbitrary waveform generator. 

We next turn on a 1D optical tweezer array of 37 sites with depths of $\approx k_B \times 1.3\,\text{mK}$ and stochastically load single CaF molecules into the sites. Due to light-assisted collisional loss, each site is either empty or has exactly one molecule. 
These molecules are then sorted into low-defect 1D arrays that are uniformly spaced. Details can be found in~\cite{Holland2023Entanglement}. In short, we determine each tweezer's occupation through non-destructive fluorescence imaging via $\Lambda$-imaging~\cite{Cheuk2018Lambda,Anderegg2019Tweezer,Holland2023bichromatic,Holland2023Entanglement}. The emitted fluorescence, collected through the objective, is imaged onto an EMCCD camera. A thresholding procedure on detected camera counts is used to classify tweezer sites as occupied or empty. This occupation information is then used to rearrange the loaded tweezer sites into a defect-free 1D array of $\mathcal{N}=8$ sites uniformly spaced at $4.2\,\mu\text{m}$ where interactions are negligible.

After sorting, we prepare the product state $|{\downarrow\downarrow\downarrow\downarrow\downarrow\downarrow\downarrow\downarrow}\rangle$ using an optical pumping protocol along with a series of microwave pulses~\cite{Holland2025Erasure}. A measurement-enhanced state preparation protocol~\cite{Holland2025Erasure} allows internal state errors to be detected, and subsequently removed via post-selection. These post-selected arrays have negligible internal state errors ($\approx 0.005$) and an overall fidelity of $f=0.934(5)$ largely due to vacant sites.

Following state preparation, the molecules are moved to have a nearest-neighbor spacing of $2.1\,\mu\text{m}$, where they can have significant dipolar interactions. We additionally lower the tweezer depth to $\approx k_B \times 130\,\mu\text{K}$ for maximal rotational coherence. For work using XX interactions, we suppress decohering fields using an XY8 dynamical decoupling sequence~\cite{Holland2023Entanglement} composed of periodic microwave pulses applied to the $|{\uparrow}\rangle$-$|{\downarrow}\rangle$ transition at 20.5\,GHz. For work using XXZ interactions with non-zero Ising strength $\Delta$, we implement Floquet Hamiltonian engineering using a modified XY8 sequence~\cite{Lu2026SpinChain}. We use the timing between microwave pulses to tune $\Delta$. After the desired squeezing duration and any global rotations used to orient the squeezed state, we turn off interactions by rapidly transferring molecules into the non-interacting states ($|{\downarrow}\rangle=|{1}\rangle$ and $|{0}\rangle$). 

Finally, we obtain full information of the eight tweezer sites (occupation and molecular state) by using a destructive spin-resolved fluorescence imaging (SRI) protocol~\cite{Lu2026SpinChain}. This protocol allows site-resolved detection of the populations in the $|{\uparrow}\rangle$ and $|{\downarrow}\rangle$ states every experimental run, and therefore allows us to fully determine whether a tweezer site has a $|{\uparrow}\rangle$ molecule, a $|{\downarrow}\rangle$ molecule, or is empty ($|{\emptyset}\rangle$). 

For work exploring storage of squeezed states in the non-interacting hyperfine states $|{1}\rangle$-$|{0}\rangle$, we use a transduction procedure with a 2-photon Raman pulse and a microwave pulse that preserves coherence during transduction. In addition, to suppress decohering fields, we apply XY8 dynamical decoupling using shaped 2-photon Raman pulses resonant with the hyperfine transition. Details can be found in~\cite{Holland2025Erasure}. 

For all measurements, we convert fluorescence images obtained via SRI into binarized data. Specifically, for every run, which is composed of multiple fluorescence images, for each image, a tweezer site is classified as bright or dark according to a thresholding procedure on the detected camera counts. These digitized datasets are then used to post-select for arrays largely free of vacancies and internal state errors. They are also used to extract the state of each tweezer site prior to SRI. Depending on the reported quantity, measurement correction is applied. Our measurement correction procedure corrects for both classification errors in the thresholding procedure and also for process errors in SRI. The parameters for measurement correction are characterized independently.

Further experimental details for Floquet engineering, internal state preparation, switching-off of interactions, SRI protocol, and measurement correction can be found in the following sections.

\section{Experimental Procedure Details} 
\subsection{Floquet Engineering of XXZ Models}
In this section, we describe how tunable $1/r^3$ XXZ models in the rotational encoding $|{\uparrow}\rangle$-$|{\downarrow}\rangle$ are generated via Floquet Hamiltonian engineering. More details, including engineering of XYZ models not explored in this work, can be found in Ref.~\cite{Lu2026SpinChain}. In brief, we utilize two pulse sequences in our work. For XX squeezing, we use a standard XY8 sequence shown in Fig.~\ref{fig:theory_sequences}(a), where a series of $\pi$ pulses about $x$ and $y$ are applied. This provides robustness to microwave detuning errors, but preserves the native dipolar spin exchange interactions. For XXZ squeezing, we use a modified XY8-XYZ sequence where each $\pi$-pulse in XY8 is split into two $\pi/2$ pulses separated by free evolution time, as shown in Fig~\ref{fig:theory_sequences}(b). By adjusting the three free-evolution times ($\{\tau_1, \tau_2, \tau_3\}$), $1/r^3$ XYZ interactions of the form
\begin{eqnarray}
	\hat{H}_{\text{XYZ}} &=& \sum_{i<j} \frac{\hbar}{|i-j|^3} \left[ J_x \hat{S}_i^x\hat{S}_j^x +J_y \hat{S}_i^y\hat{S}_j^y +J_z \hat{S}_i^z \hat{S}_j^{z} \right] \nonumber \\
	&=& \sum_{i<j} \frac{\hbar J}{|i-j|^3} \left[ \frac{1}{2}(\hat{S}_i^+\hat{S}_j^- + \hat{S}_i^-\hat{S}_j^+) + \right.\nonumber \\
	& & \left. \frac{\gamma}{2}(\hat{S}_i^+\hat{S}_j^+ + \hat{S}_i^-\hat{S}_j^{-})+\Delta \hat{S}_i^z \hat{S}_j^z\right]. 
	\label{eq:XYZ} 
\end{eqnarray}
Our work explores XXZ models where $\gamma=0$. The Floquet engineering pulses limit the accessible range of Ising strengths to $\Delta\in[0,2]$. Table~\ref{tab:floquet_sequences} shows the timing parameters used, and the measured effective depolarization ($T_1$) and decoherence times ($T_2$). To achieve long coherence times, we operate at a pseudo-magic condition where the tweezer has depth $V_0 \approx k_B \times 130\,\mu\text{K}$ and the magnetic field is set at $4.4\,\text{G}$ orthogonal to the propagation direction of the  laser that generates the tweezers. To minimize interaction disorder arising from finite motional temperature, the magnetic field is oriented orthogonal to the 1D tweezer array axis and the tweezer polarization is aligned along  the array direction.

\begin{table*}
\centering
\begin{ruledtabular}
\begin{tabular}{llcccccccc}
$\Delta$ & $\Delta_{\rm exp}$ & $\gamma$ & Sequence Type & $\tau_{1}\,(\mu{\rm s})$ & $\tau_{3}\,(\mu{\rm s})$ & $T\,(\mu{\rm s})$ &$T_1$ (ms) & $T_2$ (ms) \\
\hline
0 &0.012(11) & 0 & XY8 & 400 & N.A. & 3226.4 & 5800(900)& 1060(200)\\
0.33 & 0.361(8) & 0 & XY8-XYZ & 66.7 & 24.7 & 784 &890(10) & 810(50)\\
0.67 & 0.686(10) & 0 & XY8-XYZ & 46.7 & 46.7 & 800 &1020(60) & 780(80)\\
0.80 & 0.860(8) & 0 & XY8-XYZ & 32.7 & 44.7 & 800 &920(70) & 1020(130)\\
1.33 & 1.329(11) & 0 & XY8-XYZ & 8.7 & 44.7 & 480 & 930(140) & 389(23)\\
1.8& 1.741(19) & 0 & XY8-XYZ & 2 & 91.4 & 800 & 920(170)& 429(18)\\
\end{tabular}
\end{ruledtabular}
\caption{\label{tab:floquet_sequences} Floquet microwave pulse sequences and performance. Shown are pulse parameters used to realize XX and XXZ squeezing interactions explored in this work. Here $\tau_1$ and $\tau_3$ are the timing parameters in the pulse sequence. These free-evolution times are measured between the edges of adjacent pulses. $T$ is the total Floquet period. $T_1$ and $T_2$ show $1/e$ depolarization times and decoherence $1/e$ times with blackbody leakage rates subtracted, respectively. For all sequences, the pulse width is $t_p=3.3\,\mu{\rm s}$. $\Delta$ denotes the target Ising strength, used throughout the main text, while $\Delta_{\rm exp}$ is the experimentally inferred $\Delta$ obtained from fitting two-molecule dynamics (see~\cite{Lu2026SpinChain} for detailed procedure).}
\end{table*}

\subsection{Internal State Preparation}
In this section, we describe in detail our internal state preparation procedure. Following the first non-destructive image that determines the tweezer occupation following stochastic loading, molecules are optically pumped into $|{D'}\rangle=|N=1, F=2, m_F=-2\rangle$. Next, the tweezer depth is reduced and the magnetic field is oriented for optimal population transfer on the $|{D'}\rangle\rightarrow|{-}\rangle=|N=0, F=1, m_F=-1\rangle$ transition. The magnetic field (4.4\,G) is then rotated to be perpendicular to the array axis and tweezer polarization for maximal rotational coherence.  A microwave pulse transfers molecules from $|{-}\rangle$ to $|{\uparrow}\rangle=|N=1, F=0, m_F=0\rangle$ and a second pulse transfers molecules from $|{\uparrow}\rangle$ to $|{\downarrow}\rangle=|N=0, F=1, m_F=0\rangle$. Following this transfer, the molecules are in $|{\downarrow}\rangle$. We wait $80\,\text{ms}$ for the magnetic field to settle, during which time, $v=1$ repumping light is applied. This light addresses the $X^2\Sigma(v=1, N=1)\rightarrow A^2\Pi_{1/2}(v=0,J=1/2,+)$ transition, and converts population leakage errors arising from blackbody radiation into detectable population in $X^2\Sigma(v=0,N=1)$. At the end of this $80\,\text{ms}$ period, we apply a rapid resonant imaging protocol which allows flagging leakage errors and earlier errors in both optical pumping and microwave transfer~\cite{Holland2025Erasure}. For all experiments, we post-select on shots where no errors are detected in this image, thereby suppressing internal state preparation errors.

\subsection{Switching off Interactions Post-Squeezing}
To rapidly turn off dipolar interactions following squeezing, we transfer population out of the interacting manifold into non-interacting hyperfine states. In particular, we apply a $\pi$-pulse to transfer $|{\uparrow}\rangle$ molecules to $|{0}\rangle = |N=0, F=0, m_F=0\rangle$. This is nominally a forbidden microwave transition, but becomes allowed at the finite magnetic fields that we work at. Molecules are then either in $|{\downarrow}\rangle$ or $|{0}\rangle$. Due to dipolar selection rules, $|\downarrow\rangle$ and $|{0}\rangle$ have negligible interactions. In detail, following squeezing and after orienting the squeezed state for readout, we transfer the population from $|{\uparrow}\rangle \rightarrow |{0}\rangle$ using a disorder-robust $90^\circ$--$180^\circ$--$90^\circ$ composite pulse sequence. The transfer pulse sequence has a duration of $\approx 400\,\mu\mathrm{s}$, which is much faster than the interaction timescale set by the dipolar coupling strength $J \approx h \times 33\,\mathrm{Hz}$. Following the transfer, the optical tweezers are separated to the imaging spacing of $4.2\,\mu\mathrm{m}$, which further strongly suppresses dipolar interactions.

\subsection{Spin-Resolved Imaging (SRI)}\label{ssec:SRIProt}
In this section, we describe a spin-resolved imaging (SRI) protocol~\cite{Lu2026SpinChain} that distinguishes between the three possible occupations of each optical tweezer: occupied by a molecule in $|\uparrow\rangle$, occupied by a molecule in $|\downarrow\rangle$, or empty ($|\emptyset\rangle$). This capability is essential for post-selecting on defect-free molecular arrays.

The protocol relies on shelving into and de-shelving from the $X^2\Sigma(v=0,N=3)$ rotational manifold, which is off-resonant to imaging light and therefore does not fluoresce during imaging. To readout the population of two states, we first shelve one of the spin states and destructively detect population in the remaining spin state. Subsequently, the shelved molecules are de-shelved back into detectable/bright $X^2\Sigma(v=0,N=1)$ states for a second image. Both spin populations can therefore be measured.

In detail, this is accomplished with two rotational pumping lasers. The first laser, denoted $N12$, is used for shelving into $N=3$. It addresses the $X^2\Sigma(v=0,N=1)\rightarrow B^2\Sigma(v=0,N=2,J=3/2)$ transition. Molecules in $X^2\Sigma(v=0,N=1)$ are rapidly pumped into the $X^2\Sigma(v=0,N=3)$ manifold after scattering only a few photons. The second laser, denoted $N32$, is used to de-shelve molecules from $N=3$ back into the detectable $N=1$ manifold. It addresses the $X^2\Sigma(v=0,N=3)\rightarrow B^2\Sigma(v=0,N=2,J=5/2)$ transition, thereby pumping molecules back into $X^2\Sigma(v=0,N=1)$. Because of highly diagonal Franck-Condon factors present in the $X-B$ transition, vibrational leakage to $v=1$ states due to optical pumping is minimal ($10^{-3}$ level).

\begin{figure*}[t]
	{\includegraphics[width=\linewidth]{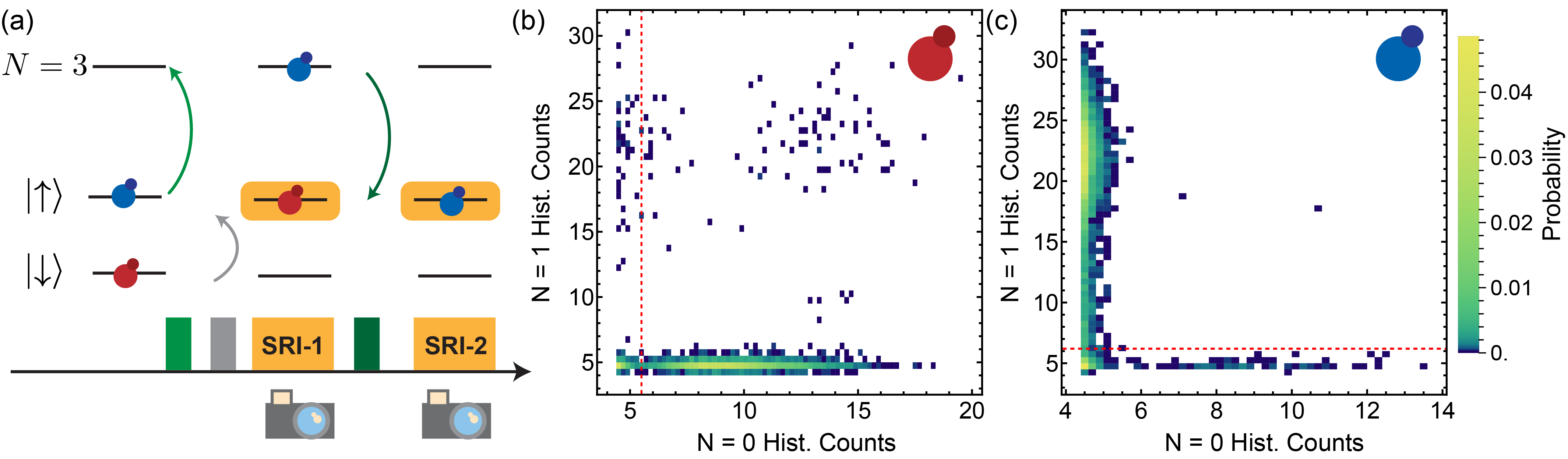}}
	\caption[SRI Sequence and Histograms]{\label{fig:sri} SRI Sequence and Histograms: (a) SRI begins with molecules in the rotational encoding. Application of $N12$ and a MW $\pi$-pulse transfers $|{\uparrow}\rangle\rightarrow N=3$ and $|{\downarrow}\rangle\rightarrow |{\uparrow}\rangle$. SRI-1 is applied to identify molecules originally in $|{\downarrow}\rangle$. Next, $N32$ is applied to return $N=3 \rightarrow N=1$ before the SRI-2 image, which identifies molecules originally in $|{\uparrow}\rangle$. (b,c) Fluorescence histograms from the two images with molecules primarily initialized in either $|{\downarrow}\rangle$ (b) or $|{\uparrow}\rangle$ (c) and detection thresholds (red dashed).}
\end{figure*}

In our work, we use the following procedure. Before detection, molecules are in the non-interacting states $|{\downarrow}\rangle=|{1}\rangle$ and $|{0}\rangle$ (after interaction switch off, or when for extended storage), and the tweezers are separated into the large distance of $4.2\,\mu\mathrm{m}$ where interactions are negligible even in interacting states. Our SRI protocol is as follows (Fig.~\ref{fig:sri}(a)):
\begin{enumerate}
	\item A microwave pulse transfers $|{0}\rangle$ molecules to $|{\uparrow}\rangle$.
	\item The tweezer depth is increased from the interaction depth $V_0$ to $7V_0$ to improve survival during rotational shelving.
	\item $N12$ light is applied to shelve molecules from $|{\uparrow}\rangle$ to the $N=3$ manifold. $0.2\,\text{ms}$ of shelving light is applied, much longer than the measured $\sim20\,\mu\text{s}$ $1/e$ shelving timescale.
	\item The tweezer depth is ramped back down to $V_0$.
	\item A microwave $\pi$-pulse transfers $|{\downarrow}\rangle$ molecules to $|{\uparrow}\rangle$.
	\item The tweezer depth is increased to the full imaging depth $\approx 10V_0$ and the magnetic field is set to zero.
	\item Molecules in $N=1$ are imaged (SRI-1). This is done using $4\,\mathrm{ms}$ of $\Lambda$-imaging followed by $7\,\mathrm{ms}$ of rapid resonant imaging~\cite{Holland2025Erasure}. Bright sites predominantly correspond to sites with molecules initially in $|{\downarrow}\rangle$. Molecules that fluoresce in this step are largely lost.
	\item $\sim10\,\mathrm{ms}$ of $N32$ light is applied to pump (de-shelve) the dark $N=3$ molecules back to detectable $N=1$ states. The duration is sufficient to readout the first image (SRI-1) from the EMCCD camera.
	\item Molecules in $N=1$ are imaged (SRI-2). This is done with $30\,\mathrm{ms}$ $\Lambda$-imaging. Bright sites largely correspond to sites with molecules initially in $|{\uparrow}\rangle$.
\end{enumerate}

Because the two images use different imaging parameters and techniques, their fluorescence count histograms are different. We therefore use separate thresholds $(\theta_1,\theta_2)$ for the two images when classifying them into bright and dark sites. In the absence of loss and cross-talk, the possible outcomes $\{(00),(01),(10)\}$, where 0 and 1 denote a classified dark or bright site, correspond to initial states $\{|{\emptyset}\rangle, |{\uparrow}\rangle, |{\downarrow}\rangle\}$. We observe that $\sim 4\%$ of population survives the first image (SRI-1), leading to the additional outcome $(11)$, which we assign to $|{\downarrow}\rangle$. Fig.~\ref{fig:sri}(b,c) shows fluorescence count histograms with molecules initialized in either $|{\uparrow}\rangle$ or $|{\downarrow}\rangle$. 

\subsection{Transduction to the Non-Interacting Hyperfine States}
For storage investigations, where the coherence of the spin-squeezed state needs to be preserved, we use a different transfer sequence compared to the interaction switch-off sequence. We transduce from the rotational encoding ($|{\uparrow}\rangle-|{\downarrow}\rangle$) to the hyperfine encoding ($|{1}\rangle-|{0}\rangle$) using a 2-photon Raman pulse and a microwave pulse.  First, a two-photon Raman $\pi$-pulse transfers population from $|{\downarrow}\rangle$ to $|{0}\rangle$. Subsequently, a microwave $\pi$-pulse transfers population from $|{\uparrow}\rangle$ to $|{1}\rangle$. 

Although coherence is maintained during the transduction process, differential ac Stark shifts between the three involved states lead to additional phase shifts. These arise primarily from single-photon Stark shifts induced by the Raman beams and produce an effective $\sim 30^\circ$ rotation about the $z$ axis in the hyperfine encoding after transduction. The observed phase shift is in excellent agreement with both the independently measured differential Stark shifts within the three-level system and the known timing of the transduction pulse sequence. To compensate for this phase accumulation, we implement a virtual $S_z$ operation on the hyperfine qubit during the transduction sequence. Experimentally, this is realized by adjusting the phase of the microwave transduction $\pi$-pulse relative to the phase of the initial microwave $\pi/2$-pulse used in the full experimental sequence.

\section{Measurement Correction Details}\label{sec:SRIMC}
This section describes the characterization and correction of errors arising from our measurement procedure with spin-resolved imaging (SRI). We also describe how post-selection is performed, and how measurement-corrected observables are extracted from our measurements. In our experiment, measurement errors fall into two categories: 1) image classification errors and 2) SRI process errors.
\subsection{Image Classification Errors}
Image classification errors are errors that occur during binarization of fluorescence counts from a site, i.e. classifying whether a site is dark or bright based on a threshold. To determine error rates, molecules are first identified using a non-destructive image, after which either SRI-1 or SRI-2 is applied. This allows extraction of $\epsilon_{10}(\theta)$ and $\epsilon_{01}(\theta)$ for each image. From these, we construct the matrix
\begin{equation}
	C_{4\times4} = 
	\underbrace{\begin{pmatrix}
			1 - \varepsilon_{01}^{(1)} & \varepsilon_{01}^{(1)} \\
			\varepsilon_{10}^{(1)} & 1 - \varepsilon_{10}^{(1)}
	\end{pmatrix}}_{\text{SRI-1}}
	\otimes
	\underbrace{\begin{pmatrix}
			1 - \varepsilon_{01}^{(2)} & \varepsilon_{01}^{(2)} \\
			\varepsilon_{10}^{(2)} & 1 - \varepsilon_{10}^{(2)}
	\end{pmatrix}}_{\text{SRI-2}}.
	\label{eq:kron}
\end{equation}
This matrix quantifies identification errors. It transforms the true occupation probability during the two images ($p_{\text{img}}$) into measured outcome probabilities after binarization ($p_\text{m}$). Explicitly, we have $p_\text{m}=C_{4\times4}\cdot p_{\text{img}}$.

\subsection{SRI Process Errors}
In addition to classification errors, there are process errors that occur during the SRI procedure. We consider errors of the full SRI process, including loss and cross-talk, without attempting to isolate individual contributions. We first determine the state-preparation fidelity for $|{\downarrow}\rangle$ by preparing this state, transferring it to $|{\uparrow}\rangle$, and detecting population with a $30\,\mathrm{ms}$ $\Lambda$-image. The true preparation probability is obtained via
\begin{equation}
	\begin{pmatrix} p_{\rm SP}^{\rm true} \\ 1 - p_{\rm SP}^{\rm true} \end{pmatrix}
	=
	\begin{pmatrix} 1 - \varepsilon_{10} & \varepsilon_{01} \\ \varepsilon_{10} & 1 - \varepsilon_{01} \end{pmatrix}^{-1}
	\begin{pmatrix} p_{\rm SP}^{\rm raw} \\ 1 - p_{\rm SP}^{\rm raw} \end{pmatrix}.	
\end{equation}
This is then used to construct the state-preparation matrix
\begin{equation}\label{eq:spmat}
	M_{\rm SP} =
	\begin{pmatrix}
		p_{\rm SP}^{\rm true} & 0 & 0 \\
		0 & p_{\rm SP}^{\rm true} & 0 \\
		1 - p_{\rm SP}^{\rm true} & 1 - p_{\rm SP}^{\rm true} & 1
	\end{pmatrix}.
\end{equation}
The rows of this matrix quantify how well we can prepare a molecule in $|{\uparrow}\rangle$, $|{\downarrow}\rangle$, or $|{\emptyset}\rangle$, where we assume a perfect $\pi$-pulse fidelity. (We estimate this infidelity to be at the $10^{-3}$ level, which is much smaller compared to other errors and can therefore be ignored.)
Next, we apply SRI to tweezers prepared in $|{\uparrow}\rangle$, $|{\downarrow}\rangle$, or $|{\emptyset}\rangle$. We compile the measurement outcomes as a function of SRI imaging thresholds into the matrix
\begin{equation}\label{eq:pout}
	P_{\rm out}(\theta_1, \theta_2) =
	\begin{pmatrix}
		P_{11}^{(N{=}0)} & P_{11}^{(N{=}1)} & P_{11}^{(\rm dark)} \\
		P_{10}^{(N{=}0)} & P_{10}^{(N{=}1)} & P_{10}^{(\rm dark)} \\
		P_{01}^{(N{=}0)} & P_{01}^{(N{=}1)} & P_{01}^{(\rm dark)} \\
		P_{00}^{(N{=}0)} & P_{00}^{(N{=}1)} & P_{00}^{(\rm dark)}
	\end{pmatrix}.
\end{equation}
Finally, we introduce the projection matrix,
\begin{equation}\label{eq:proj}
	\Pi =
	\begin{pmatrix}
		1 & 1 & 0 & 0 \\
		0 & 0 & 1 & 0 \\
		0 & 0 & 0 & 1
	\end{pmatrix},
\end{equation}
which conveys that, after correcting for identification errors, $(11)$ and ($10$) arise from a molecule in $N=1$ during the SRI-1 image.

From here, we can relate the inferred occupation of the tweezers following SRI to the true state preparation probabilities $M_\text{SP}$ via the SRI process matrix ($\mathcal{M}$):
\begin{equation}
	\Pi \cdot C_{4\times4}^{-1}\cdot P_\text{out}= \mathcal{M} \cdot M_\text{SP}
\end{equation}
Solving for the process matrix we obtain
\begin{equation}
	\mathcal{M}=\Pi \cdot C_{4\times4}^{-1}\cdot P_\text{out}\cdot M_\text{SP}^{-1}.
\end{equation}

Finally, the full correction matrix $\mathcal{R}$, relating a true distribution $p_\text{t}$ to a measured distribution $p_\text{m}$, is
\begin{equation}
	p_\text{t}=\underbrace{\mathcal{M}^{-1} \cdot \Pi \cdot C_{4\times4}^{-1}}_{\mathcal{R}} \cdot p_\text{m}.
\end{equation}
Written in full, we have
\begin{equation}
	\mathcal{R}=M_\text{SP}\cdot [\Pi \cdot C_{4\times4}^{-1} \cdot P_\text{out}]^{-1} \cdot \Pi \cdot C_{4\times4}^{-1}
\end{equation}
where $M_\text{SP}$ is the $3\times3$ state-preparation matrix (Eq.~\ref{eq:spmat}), $\Pi$ is the $3\times4$ projection matrix (Eq.~\ref{eq:proj}), $C_{4\times4}^{-1}$ is the inverse image-identification error matrix (Eq.~\ref{eq:kron}), and $P_\text{out}$ is the $4\times3$ measured probability matrix (Eq.~\ref{eq:pout}).

\subsection{Post-selection and Expectation Values}
The single-tweezer correction matrix generalizes to an array of $N$ tweezers as $\mathcal{R}_N=\mathcal{R}^{\otimes N}$. While this matrix becomes large (for $N=8$, $\mathcal{R}^{\otimes N}$ is $3^8\times4^8$), it only needs to be stored temporarily during the correction step.

To perform measurement correction and post-selection:
\begin{enumerate}
	\item Apply $\mathcal{R}^{\otimes N}$ to the measured outcome vector $P_\text{meas}$ (dimension $4^N$), yielding the occupation probability vector $P_\text{occ}$ (dimension $3^N$).
	\item Post-select by setting all entries of $P_\text{occ}$ corresponding to hole configurations to zero.
	\item Normalize the resulting vector's sum to 1.
\end{enumerate}
Before normalization, this vector can be written as
\begin{align}
	P_\text{occ,PS}&=\text{diag}\{1,1,0\}^{\otimes N}\cdot\mathcal{R}^{\otimes N}\cdot P_\text{meas}\nonumber\\
	&=[\text{diag}\{1,1,0\}\cdot\mathcal{R}]^{\otimes N}\cdot P_\text{meas}.
\end{align}
Since $\text{diag}\{1,1,0\}\cdot\mathcal{R}$ contains a row of zeros, the correction matrix reduces to dimension $2\times4$, significantly reducing computational overhead.

Expectation values and uncertainties are computed via bootstrap resampling. Starting from the binarized measurement outcomes, we generate resampled datasets, apply correction and post-selection, and compute expectation values $\{\langle \mathcal{O} \rangle\}_i$. Reported values correspond to the mean and standard deviation over bootstrap samples.

The results in the paper are obtained with arrays post-selected for zero vacancies, except for the hyperfine storage results. For hyperfine transduction, we do not apply post-selection, so that the comparison between spin-squeezed and coherent spin state is as fair as possible (same data rate). For plots where we report uncorrected results with post-selection applied, post-selection corresponds to requiring at least one bright image per tweezer during SRI.

\section{Bipartite Entanglement and EPR Steering Analysis Details}
In this section, we provide details regarding the witnesses for bipartite entanglement and EPR steering.

In our work, we use the witnesses for bipartite entanglement and EPR steering from Ref.~\cite{Fadel2018BECEPR}. 
For bipartite entanglement, the separability witness in its most general form is given by
\begin{align}
	\mathcal{W}
	=
	\frac{
		4\,\mathrm{Var}\!\left( g_z \hat{S}_z^A + \hat{S}_z^B \right)
		\,\mathrm{Var}\!\left( g_x \hat{S}_x^A + \hat{S}_x^B \right)
	}{
		\left( |g_z g_x| \, \left( |\langle \hat{S}_y^A \rangle| + |\langle \hat{S}_y^B \rangle| \right) \right)^2
	},
\end{align}
where $\hat{S}_z^A=\sum_{i\in A}\hat{S}_z^i$ (likewise for $\hat{S}_z^B$), and $g_z$ and $g_x$ are free-parameters that can be chosen to minimize $\mathcal{W}$. The separability witness $\mathcal{W}$ in the main text is obtained when we fix these free parameters to $g_z=1$ and $g_x=-1$, giving the specific expression of $\mathcal{W}=4\text{Var}[\hat{S}_z]\text{Var}[\hat{S}_x^a]/|\langle \hat{S}_y \rangle|^2$. We apply this witness to the squeezed state oriented with its minimum variance along $z$. 

For EPR steering, the witness $\mathcal{S}\mathcal{S}_{A\rightarrow B}$ takes the form
\begin{align}
	\mathcal{S}_{A\rightarrow B}=\frac{
		\mathrm{Var}\!\left[ \left( \sum_{i} g_z^i \hat{S}_z^i \right) + \hat{S}_z^B \right]
		\,
		\mathrm{Var}\!\left[ \left( \sum_{i} g_x^i \hat{S}_x^i \right) + \hat{S}_x^B \right]
	}{
		\left| \left\langle \hat{S}_y^B \right\rangle \right|^2 / 4
	},
\end{align}
where $i\in A$, and $g_\beta^i$ are free parameters that can be chosen to minimize $\mathcal{S}_{A\rightarrow B}$. For the EPR witnesses presented in the main text, we use values for $g_\beta^i$ theoretically determined for disorder-free squeezing. Table.~\ref{tab:EPRgs} provides the weights $g_\beta^i$ used for the EPR witnesses presented in the main text. In detail, we determine these weights by simulating disorder-free $1/r^3$ XXZ squeezed states at $t^*$ for each Ising strength $\Delta$. The eight weights are then numerically optimized to minimize $\mathcal{S}_{A \rightarrow B}$. Due to the reflection symmetry of the 8-site chain and the even/odd partition into the A/B subsystems, weights for $\mathcal{S}_{B \rightarrow A}$ are related by a mirror symmetry on the indices.

\begin{table}[h!]
	\begin{center}
		\begin{tabular}{ c|c|c|c|c|c|c|c|c }
			$\Delta$ & $g^1_z$ & $g^3_z$ & $g^5_z$ & $g^7_z$ & $g^1_x$ & $g^3_x$ & $g^5_x$ & $g^7_x$\\
			\hline
			0    & 0.445 & 0.727 & 0.733 & 0.831 & -0.414 & -0.751 & -0.717 & -0.898 \\
			0.33 & 0.450 & 0.632 & 0.663 & 0.744 & -0.414 & -0.637 & -0.659 & -0.844 \\
			0.5  & 0.499 & 0.595 & 0.658 & 0.696 & -0.471 & -0.539 & -0.734 & -0.850 \\
			0.67 & 0.586 & 0.607 & 0.653 & 0.669 & -0.528 & -0.593 & -0.667 & -0.919 \\
			0.8  & 0.668 & 0.636 & 0.621 & 0.682 & -0.692 & -0.601 & -0.641 & -0.857 \\
			1.33 & 0.446 & 0.516 & 0.604 & 0.518 & -0.337 & -0.500 & -0.649 & -0.688 \\
			1.8  & 0.374 & 0.506 & 0.567 & 0.568 & -0.341 & -0.453 & -0.589 & -0.637 \\
		\end{tabular}
	\end{center}
	\caption{\label{tab:EPRgs}EPR Witness Parameters vs $\Delta$: The $g_\beta^i$ that minimize $\mathcal{S}_{A \to B}$ are obtained from noiseless disorder-free simulations. The mirror-symmetry of the array geometry implies that optimal parameters for $\mathcal{S}_{B \to A}$ can be obtained by substituting spatial indices according to $\{1,3,5,7\}\rightarrow\{8,6,4,2\}$.}
\end{table}

\section{Modeling Decoherence and Disorder}
In the following, we discuss our noise model and its implementation. We examine the individual noise source in Secs.~\ref{ssec:theory_tweezer}, \ref{ssec:theory_microwave}, and \ref{ssec:theory_noninteracting}. We then explain how we fit the parameters of our model in Sec.~\ref{ssec:theory_fitting}. Finally, we discuss the numerical implementation in Sec.~\ref{ssec:numerical_details}.

\subsection{Tweezer Potential and Molecular Temperature} \label{ssec:theory_tweezer}

\begin{table}[]
    \centering
    \begin{tabular}{l|l}
        Parameter & Value \\
        \hline
        $a$                             & 2100\,nm \\
        $\omega_x/(2\pi)$               & 61.3\,kHz \\
        $\omega_y/(2\pi)$               & 67.4\,kHz \\
        $\omega_z/(2\pi)$               & 15\,kHz \\
        $\delta_x/(2\pi)$               & 20\,Hz \\
        $\delta_y/(2\pi)$               & 15\,Hz \\
        $\delta_z/(2\pi)$               & 3\,Hz \\
        $\bar n_x$                      & 1.4 \\
        $\bar n_y$                      & 1.4 \\
        $\bar n_z$                      & 26 \\
        $a_{\mathrm{ho},x}$             & 37.4\,nm \\
        $a_{\mathrm{ho},y}$             & 35.7\,nm \\
        $a_{\mathrm{ho},z}$             & 75.6\,nm \\
    \end{tabular}
    \caption{Tweezer Trapping Parameters used for Simulations.}
    \label{tab:trapparameters}
\end{table}

\subsubsection*{Single Molecule Eigenstates}
For the ideal $1/r^3$ XXZ model, molecules are perfectly localized at the center of each tweezer. In reality, due to finite motional temperature, the molecules move around in their respective tweezer. To capture this motion, we model each tweezer as three uncoupled harmonic oscillators along the $x$, $y$, and $z$ directions.
We model the internal degrees of freedom as two-level systems, such that a single molecule in a tweezer is described by the effective Hamiltonian
\begin{align}
    \hat H_{\rm eff} =&
    \hbar \omega \hat s^{(z)} +
    \hbar \omega_x \hat a_x^\dagger \hat a_x +
    \hbar \omega_y \hat a_y^\dagger \hat a_y +
    \hbar \omega_z \hat a_z^\dagger \hat a_z 
    \nonumber \\
    & + \hbar \qty(\delta_{x,\rm th} \hat a_x^\dagger \hat a_x + \delta_{y,\rm th} \hat a_y^\dagger \hat a_y + \delta_{z,\rm th} \hat a_z^\dagger \hat a_z) \hat s^{(z)}
    \, .
    \label{eq:tweezer_model}
\end{align}
where $\hat a_x$, $\hat a_y$, and $\hat a_z$ annihilate motional excitations along the $x$, $y$, and $z$ directions. The parameters $\delta_{\alpha, \rm th}$ capture the fact that the tweezer depths (and hence trapping frequency) depend on the internal molecular state~\cite{caldwell2020sideband,Lu2024RSC}. Since the tweezers are almost magic, we have $\delta_{\alpha, \rm th} \ll \omega_\alpha$.

We determine the parameters of Eq.~\eqref{eq:tweezer_model} from a Gaussian beam model of the tweezer potential with known molecular polarizabilities~\cite{caldwell2020sideband,Lu2024RSC}. We first project the tweezer light field onto a grid of size $L_x \times L_y \times L_z$, centered around the point of peak intensity. At each grid-point $\vec r$, we diagonalize the rotation-hyperfine Hamiltonian of CaF in the tweezer AC electric field $\vec E(\vec r, \omega)$
\begin{align}
    \hat H_{\rm mol}(\vec r) &=
    -\vec E^\dagger (\vec r, \omega) \hat \alpha(\omega) \vec E(\vec r, \omega)
    - \vec \mu_{\rm mol} \vec B
    + \hat H_{\rm rot} + \hat H_{\rm hf}
    \, ,
\end{align}
where $\hat \alpha(\omega)$ is the tensor polarizability of the molecule. The local tweezer field $\vec E$ follows a Gaussian beam model with linearly polarized light in the far field orthogonal to the applied magnetic field of $\vec B = 4.4$ G $\times \hat e_y$, which is orthogonal to the tweezer array axis  direction.

In practice, we diagonalize $\hat H_{\rm mol}$ for states with rotational angular momentum $N \leq 1$, which is sufficient to reach convergence in all parameters.
We find eigenstates $\ket{\lambda_l(\vec r)}$ with $\hat H_{\rm mol}(\vec r) \ket{\lambda_l(\vec r)} = V_l(\vec r) \ket{\lambda_l(\vec r)}$.
These eigenstates depend only weakly on the position $\vec r$.
We can then generally write single-molecule states in this basis as $\ket{\phi_{n,l}} = \int d\vec r \phi_n^{(l)}(\vec r) \ket{\lambda_l(\vec r)}$, where $\phi_n^{(l)}(\vec r)$ are motional eigenstates for molecules in state $l$.

Next, we write the single-molecule Hamiltonian $\hat H_{0} = - \hbar^2\nabla^2/2m + \sum_l V_l(\vec r) \ket{\lambda_l(\vec r)} \bra{\lambda_l(\vec r)}$ in this new basis. By construction, $V_l$ does not couple $\ket{\phi_{n,l}}$ and $\ket{\phi_{n',l'}}$ for $l \neq l'$. Furthermore, $\ket{\lambda_l(\vec r)}$ only slowly varies with $\vec r$ such that $\hbar^2 \bra{\lambda_l(\vec r)} \nabla^2 \ket{\lambda_{l'}(\vec r)}/2m \ll V_l(\vec r) - V_{l'}(\vec r)$. We can therefore ignore the off-diagonal couplings and approximate $\hat H_0 \approx \hat H_{0,l}$ with $\hat H_{0,l} = [\hbar^2\nabla^2/2m + V_l(\vec r)] \ket{\lambda_l(\vec r)} \bra{\lambda_l(\vec r)}$.
This approximation is reminiscent of the Born-Oppenheimer approximation, but now acts on external motion and internal states. We can then diagonalize the spatial component independently for each $l$ to find the eigenstates and eigenenergies $\hat H_{0,l} \ket{\phi_{n,l}} = E_{n,l} \ket{\phi_{n,l}}$.

Following Ref.~\cite{colbert1992novel}, we compute the lowest eigenstates and eigenenergies of $\hat H_{0,l}$ for the two spin states $\downarrow$ ($l=3$) and $\uparrow$ ($l=8$).
This gives us the full 3D motional eigenstates.
To disentangle the $x$, $y$, and $z$ components, we take three 1D cuts of the Hamiltonian.
We define $\hat H_{0,l}^{(x)}$ and $E_{n,l}^{(x)}$ by fixing $y=z=0$ to compute $\omega_x = (E_{3,1}^{(x)} - E_{3,0}^{(x)})/\hbar$ and $\delta_{x,\rm th} = (E_{8,1}^{(x)} - E_{8,0}^{(x)})/\hbar - \omega_x $.
We analogously compute $\omega_y$ and $\delta_{y,\rm th}$ for $x=z=0$, and $\omega_z$ and $\delta_{z,\rm th}$ for $x=y=0$. 
We call these disentangled eigenstates $\ket{\phi_{\vec n, l}}$, where $\vec n = (n_x, n_y, n_z)^T$ label the three spatial components and $l=\uparrow/\downarrow$ is the pseudo-spin.
The resulting parameters are given in Tab.~\ref{tab:trapparameters}.
We confirm that the low-energy spectrum matches that of the full 3D model. 

\subsubsection*{Interactions}

In general, dipolar exchange interactions are given by
\begin{align}
    \hat H_{\rm dd} =
    \frac{d^2}{4\pi\epsilon_0}
    &\int d^3\vec r_i d^3\vec r_j \frac{1 - 3\frac{\qty[\hat e_y \cdot (\vec r_i - \vec r_j)]^2}{\abs{\vec r_i - \vec r_j}^2}}{\abs{\vec r_i - \vec r_j}^3}
    \nonumber \\
    &\quad \times
    \hat \psi^\dagger_\uparrow(\vec r_i) \hat \psi^\dagger_\downarrow(\vec r_j) \hat \psi_\uparrow(\vec r_j) \hat \psi_\downarrow(\vec r_i)
    \, ,
\end{align} Here $\hat \psi^\dagger_\sigma(\vec r_i) $ ($\hat \psi_\sigma(\vec r_i) $) are field  operators that create (annihilate) a molecule at position $\vec r_i$ and state $\sigma=\uparrow,\downarrow$. $\hat e_y$ denotes the quantization axis, and $d$ is the transition dipole moment.

Let $\ket{\phi_{n,l}(m)}$ label the motional state of molecule $m$, with a tweezer centered at $ma\hat e_x$ and tweezer spacing $a$.

The matrix elements $\bra{\phi_{\vec n_1',\uparrow}(0),\phi_{\vec n_2',\downarrow}(m)} \hat H_{\rm dd} \ket{\phi_{\vec n_1,\downarrow}(0),\phi_{\vec n_2,\uparrow}(m)} \ll E_{\vec n_1,\downarrow} + E_{\vec n_2,\uparrow} - E_{\vec n_1',\uparrow} - E_{\vec n_2',\downarrow} $ for most combinations, except for $\vec n_1 = \vec n_1'$ and $\vec n_2 = \vec n_2'$, and for $\vec n_1 = \vec n_2'$ and $\vec n_2 = \vec n_1'$.
Crucially, for $\vec n_1 = \vec n_2'$ and $\vec n_2 = \vec n_1'$, $\bra{\phi_{\vec n_1',\uparrow}(0),\phi_{\vec n_2',\downarrow}(m)} \hat H_{\rm dd} \ket{\phi_{\vec n_1,\downarrow}(0),\phi_{\vec n_2,\uparrow}(m)} \ll J$ typically does not contribute to the dynamics on experimental time scales. Therefore, we only keep the terms corresponding to $\vec n_1 = \vec n_1'$ and $\vec n_2 = \vec n_2'$, and $\hat H_{\rm dd}$ reduces to a spin model
\begin{align}
    \hat H_{\rm dd, eff} &= \sum_{i,j} \frac{J_{ij}(\vec n^{(i)},\vec n^{(j)})}{2} \qty[\hat s_i^{(x)} \hat s_j^{(x)} + \hat s_i^{(y)} \hat s_j^{(y)}]
    \, .
\end{align}
The matrix elements $J_{ij}$ can then be computed by assuming a thermal distribution for $n_{\alpha, i}$, with mean occupations given in Tab.~\ref{tab:trapparameters}.
We evaluate
\begin{widetext}
\begin{align}
    J_{ij}\qty(\vec n^{(i)}, \vec n^{(j)}) = \frac{d^2}{4\pi\epsilon_0} 
       \int d^3 \vec r_i \int d^3\vec r_j
    &\abs{\varphi_{n_x^{(i)}}(x_i)}^2
    \abs{\varphi_{n_y^{(i)}}(y_i)}^2
    \abs{\varphi_{n_z^{(i)}}(z_i)}^2
    \abs{\varphi_{n_x^{(j)}}(x_j)}^2
    \abs{\varphi_{n_y^{(j)}}(y_j)}^2
    \abs{\varphi_{n_z^{(j)}}(z_j)}^2
    \nonumber \\
    \times&
    \frac{1 - 3 \qty(y_i - y_j)^2 / \qty{\qty[a(i-j) + x_i - x_j]^2 + \qty(y_i - y_j)^2 + \qty(z_i - z_j)^2}}{\qty{\qty[a(i-j) + x_i - x_j]^2 + \qty(y_i - y_j)^2 + \qty(z_i - z_j)^2}^{3/2}}
    \, .
\end{align}
\end{widetext}
Here, $\varphi_n(x)$ are the 1D harmonic oscillator eigenstates.

To simplify the numerical evaluation of this integral, we Taylor expand its coefficient:
We have $\int d\alpha \alpha \abs{\varphi_{n}(\alpha)}^2 = 0$ and $\int d\alpha \alpha^2 \abs{\varphi_{n}(\alpha)}^2 = a_{\rm ho, \alpha}^2 (2n+1)$, with $a_{\mathrm{ho},\alpha} = \sqrt{\frac{\hbar}{2m\omega_\alpha}}$.
Since $a_{\rm ho, x}^2 (2\bar n_x + 1) \approx a_{\rm ho, y}^2 (2\bar n_y + 1) \ll a^2$, we can expand in $x_i$, $x_j$, $y_i$, and $y_j$ and evaluate
\begin{widetext}
\begin{align}
    J_{ij}\qty(\vec n^{(i)}, \vec n^{(j)}) = \frac{d^2}{4\pi\epsilon_0} 
        \int d z_i dz_j
    \abs{\varphi_{n_z^{(i)}}(z_i)}^2
    &\abs{\varphi_{n_z^{(j)}}(z_j)}^2
    \left\{
    \frac{1}{\qty[a^2(i-j)^2 + (z_i-z_j)^2]^{3/2}}
    -
    \frac{9a_{\mathrm{ho},y}^2\qty(n_y^{(i)} + n_y^{(j)} + 1)}{\qty[a^2(i-j)^2 + (z_i-z_j)^2]^{5/2}}
    \right.
    \nonumber
    \\
    &+
    \left.
    \frac{12a_{\mathrm{ho},x}^2\qty(n_x^{(i)} + n_x^{(j)} + 1) \qty[a^2(i-j)^2 - \qty(z_i-z_j)^2/4]}{\qty[a^2(i-j)^2 + (z_i-z_j)^2]^{7/2}}
    \right\}
    \, .
    \label{eq:Jij_neighbor}
\end{align}
\end{widetext}

Furthermore, for sufficiently large $\abs{i - j}$, we also have $a_{\rm ho, z}^2 (2\bar n_z + 1) \ll \abs{i-j}^2a^2$, and we can expand also the $z$-coordinate. In this case, we get
\begin{widetext}
\begin{align}
    J_{ij}\qty(\vec n^{(i)}, \vec n^{(j)}) =
    \frac{d^2}{4\pi\epsilon_0 a^3\abs{i-j}^3}
    \qty[
    1
    +
    \frac{12a_{\mathrm{ho},x}^2}{a^2\abs{i-j}^2} 
    \qty(n_x^{(i)} + n_x^{(j)} + 1) -
    \frac{9a_{\mathrm{ho},y}^2}{a^2\abs{i-j}^2}
    \qty(n_y^{(i)} + n_y^{(j)} + 1)
    -
    \frac{3a_{\mathrm{ho},z}^2}{a^2\abs{i-j}^2}
    \qty(n_z^{(i)} + n_z^{(j)} + 1)
    ]
    \, .
    \label{eq:Jij_perturbative}
\end{align}
\end{widetext}
In practice, we find that this is a good approximation for $\abs{i - j} \geq 2$, but not $\abs{i - j} = 1$.
Thus, we use Eq.~\eqref{eq:Jij_neighbor} to compute $J_{ij}$ for nearest neighbor, and Eq.~\eqref{eq:Jij_perturbative} to compute $J_{ij}$ beyond nearest neighbor.

Finally, we write the Hamiltonian in the rotating frame of the microwave.
During free evolution, i.e.~while the microwaves are turned off, it reads
\begin{align}
    \hat H_{\rm mol} &=
    \sum_{i,j} \frac{J_{ij}}{2} \qty[\hat s_i^{(x)} \hat s_j^{(x)} + \hat s_i^{(y)} \hat s_j^{(y)}]
    +
    \sum_i \hbar \delta_i \hat s_i^{(z)}
    \, ,
    \label{eq:h_mol}
\end{align}
where the site disorder $\delta_i$ is determined by the motional states as 
\begin{align}
    \delta_i = \delta + 
    \sum_{\alpha = x,y,z}
    \delta_{\alpha, \rm th} n_{\alpha}^{(i)}
    \, ,
    \label{eq:delta_i}
\end{align}
with $\delta$ the detuning between microwave frequency and molecule transition frequency.

\subsubsection*{Coherent Pulse Dynamics}
During the microwave pulses, the Hamiltonian is given by $\hat H_{\rm mol} + \hat H_{\rm MW,\alpha}$, with
\begin{align}
    \hat H_{\rm MW,\alpha} &=
    \Omega \hat S^{(\alpha)} \, ,
\end{align}
with $\alpha = x,y$ and $\hat S^{(\alpha)} = \sum_i \hat s^{(\alpha)}_i$.

To evolve the states, we define unitary operators $\hat {\mathcal U}$. The pulse dynamics for a pulse of length $\tau_{\rm p}$ is described by 
\begin{align}
    \ket{\psi(t+\tau_{\rm p})} &=
    \hat{\mathcal U}_{\rm p,\alpha}\ket{\psi(t)}
    \nonumber \\
    &\equiv \exp[-\mi \tau_{\rm p} \qty(\hat H_{\rm mol} + \hat H_{\rm MW,\alpha})] \ket{\psi(t)}
    \, .
\end{align}
Here, $\tau_{\rm p}\Omega = \pi$ for the XY8 pulse sequence, and $\tau_{\rm p}\Omega = \pi/2$ for the Floquet sequence. Analogously, the free evolution for time $\tau$ between two pulses is then described by
\begin{align}
    \ket{\psi(t + \tau)} &=
    \hat {\mathcal U}_{\rm f}(\tau)\ket{\psi(t)}
    \equiv \exp(-\mi \tau \hat H_{\rm mol}) \ket{\psi(t)}
    \, .
\end{align}
We can then alternate between the free evolution and the pulses to describe the full coherent dynamics, e.g., for the XY8 sequence in Fig.~\ref{fig:theory_sequences}, we have
\begin{align}
    \ket{\psi(T)} &=
   \hat{ \mathcal U}_{\rm f}\qty(\frac{\tau}{2}) \hat{\mathcal U}_{\mathrm{p}, x} \hat{\mathcal U}_{\rm f}(\tau) \hat{ \mathcal U}_{\mathrm{p}, y} \cdots \hat{\mathcal U}_{\mathrm{p}, x} \hat{\mathcal U}_{\rm f}\qty(\frac{\tau}{2}) \ket{\psi(0)}
    \, ,
\end{align}
with $T = 8\tau_{\rm p} + 8\tau$ being the full sequence time. In addition to temperature, we also account for stochastic noise sources, as discussed below.

\subsection{Incoherent Dynamics --- Single-Particle Dephasing, Microwave Noise and Blackbody Radiation} \label{ssec:theory_microwave}
\subsubsection*{Single-Particle Dephasing}
While for purely unitary evolution, stochasticity only enters through thermal sampling of the initial state, additional noise sources also induce stochastic dynamics of the states. For single-molecule dephasing, we can describe the dynamics of the density matrix $\hat \rho$ using a Lindblad master equation $\partial_t \hat \rho = \mathcal L \hat \rho$ with
\begin{align}
    \mathcal L_{\rm f/(p,\alpha)} \hat \rho =
    & \frac{- \mi}{\hbar} \qty[\hat H_{\rm f/(p,\alpha)}, \hat \rho] 
    \nonumber \\
    & - \frac{1}{2} \sum_k \qty(\hat L_k^\dagger \hat L_k \hat \rho + \hat \rho \hat L_k^\dagger \hat L_k - 2 \hat L_k \hat \rho \hat L_k^\dagger)
    \, .
    \label{eq:master}
\end{align}
Here, $\hat H_{\rm f/p}$ is the relevant Hamiltonian, either describing free evolution with $\hat H_{\rm f} \equiv \hat H_{\rm mol}$ or pulse evolution with $\hat H_{\rm p,\alpha} \equiv \hat H_{\rm mol} + \hat H_{\rm MW,\alpha}$. $\hat L_k$ are the jump operators determined by the noise.

Single-molecule dephasing is described by local jump operators for each molecule, given by
\begin{align}
    \hat L_i = \sqrt{\gamma_{\rm deph}} \hat s^{(z)}_i \, ,
\end{align}
with dephasing rate $\gamma_{\rm deph}$. This can be caused by uncorrelated tweezer intensity fluctuations or fluctuations of the local magnetic fields.

Analogous to unitary evolution, we describe the evolution of the master equation with
\begin{align}
    \hat \rho(t+\tau) &=
    \exp(\mathcal L_{\rm f} \tau) \hat \rho(t) 
    & \text{(for free evolution)}
    \, ,
    \\
    \hat \rho(t + \tau_{\rm p}) &=
    \exp(\mathcal L_{\rm p} \tau_{\rm p}) \hat \rho(t) 
    & \text{(for pulse evolution)}
    \, .
    \label{eq:pulse_evolution}
\end{align}

\subsubsection*{Blackbody Radiation}
At long times, the effects of blackbody radiation are no longer negligible. Through blackbody radiation, molecules are slowly excited from the vibrational ground state $\nu = 0$ to the vibrational excited state $\nu = 1$ at a rate of $\gamma_{\rm bbr} = 0.3$ s$^{-1}$. During this process, the rotational quantum number $N$ changes by $\Delta N = \pm 1$, which can also change the angular momentum $\vec J = \vec L + \vec N + \vec S$, and $\vec F = \vec J + \vec I$  (for orbital electron angular momentum $\vec L$, electron spin $\vec S$, and nuclear spin $\vec I$). This change $\Delta N$ only trivially affects the subsequent dynamics and the measurements, and we thus do not resolve the precise changes in $\vec J$ and $\vec F$ in our simulations.

Instead, we label the excited states collectively as $\ket{\nu=1, \downarrow} = \ket{X,\nu = 1, N = 1, J, F, m_F}$ and $\ket{\nu=1, \uparrow} = \ket{X,\nu = 1, N = 0 \text{ or } 2, J, F, m_F}$, where $X$ labels the electronic ground state. We change the $N$-association of $\uparrow$ and $\downarrow$ compared to the $\nu = 0$ manifold, such that blackbody radiation leaves the pseudo-spin unchanged. Then, we can describe blackbody radiation by Lindblad jump operators
\begin{align}
    \hat L_{\mathrm{bbr},i,\downarrow} &=
    \sqrt{\gamma_{\rm bbr}} \ket{\nu = 1, \downarrow}\bra{\downarrow}
    \, ,
    \\
    \hat L_{\mathrm{bbr},i,\uparrow} &=
    \sqrt{\gamma_{\rm bbr}} \ket{\nu = 1, \uparrow}\bra{\uparrow}
    \, .
\end{align}

These vibrationally excited states decay back to the vibrational ground state $\nu = 0$ at a rate of $\gamma_{\rm bbr}' = 5.3$ s$^{-1}$, described by Lindblad jump operators
\begin{align}
    \hat L_{\mathrm{bbr},i,\downarrow}' &=
    \sqrt{\gamma_{\rm bbr}'} \ket{\downarrow'}\bra{\nu = 1, \downarrow}
    \, ,
    \\
    \hat L_{\mathrm{bbr},i,\uparrow}' &=
    \sqrt{\gamma_{\rm bbr}'} \ket{\uparrow'}\bra{\nu = 1, \uparrow}
    \, .
\end{align}

Here, we also collectively label the ground states $\ket{\downarrow'} = \ket{\nu = 0, N = 0 \text{ or } 2, J, F, m_F}$ and $\ket{\uparrow'} = \ket{\nu = 0, N = 1 \text{ or } 3, J, F, m_F}$.
Since only molecule pairs in the state $\ket{\uparrow}$ and $\ket{\downarrow}$ can resonantly exchange excitations, we assume that the states $\ket{\uparrow'}$ and $\ket{\downarrow'}$ do not interact with the rest of the system, since they predominantly fall into different hyperfine states. However, in our detection scheme, all molecules in the vibrational ground state $\nu = 0$ and with $N=1,3$ scatter photons during the measurement.
Moreover, in the specific setup we use for measurement, all molecules in any of the states labeled as $\ket{\uparrow'}$ are detected as $\ket{\uparrow}$ and all molecules in any of the states $\ket{\downarrow'}$ are detected as $\ket{\downarrow}$. Therefore, we assume that measurements can discriminate population in different rotational manifolds but cannot resolve the specific hyperfine state, i.e. measurements can be modeled by the following projectors $\ket{\uparrow}\bra{\uparrow} + \ket{\uparrow'}\bra{\uparrow'}$ or $\ket{\downarrow}\bra{\downarrow} + \ket{\downarrow'}\bra{\downarrow'}$.

\subsubsection*{Microwave Phase Noise}
In contrast to single-molecule dephasing and blackbody radiation, microwave phase noise is not captured by a time-independent Lindblad master equation. Instead, we describe the time-correlated microwave phase noise by applying the unitary  $\ket{\psi'}(t) = \hat{\mathcal U}_z(r_t)\ket{\psi}$ each time before a pulse is applied, with
\begin{align}
   \hat{ \mathcal U}_z(r_t) &= \exp(-\mi r_t \Delta \phi \hat S^{(z)})
    \, ,
    \nonumber \\
    \hat \rho &\rightarrow
    \hat{\mathcal U}_z(r_t) \hat \rho \, \hat{\mathcal U}^\dagger_z(r_t)
    \label{eq:theory_uz_definition}
\end{align}
Here, $\Delta \phi$ is the noise amplitude and $r_t$ is a Gaussian random variable with $\overline{r_t} = 0$ and $\overline{r_t r_{t'}} = \delta_{tt'}$. In particular, to capture this noise, we need to modify Eq.~\eqref{eq:pulse_evolution} to
\begin{align}
    \hat \rho(t + \tau_{\rm p})
    &=
    \exp(\mathcal L_{\rm p} \tau_{\rm p}) \qty[\hat{\mathcal U}_{z}(r_t) \hat \rho(t) \hat{\mathcal U}_z^\dagger(r_t)]
    \, .
    \label{eq:theory_uz_densitymatrix}
\end{align}
The microwave phase noise can be equivalently described by the time-dependent Hamiltonian
\begin{align}
    \hat H_{\rm MWnoise}(t) &=
    h(t) \hat S^{(z)}
    \, .
\end{align}
Here, $h(t)$ is a random variable with mean $\overline{h(t)} = 0$ and correlation $\overline{h(t)h(t')} = S(\abs{t-t'})$.
This Hamiltonian commutes with $\hat H_{\rm mol}$. Consequently, we can decompose the unitary describing evolution under $\hat H_{\rm mol} + \hat H_{\rm MWnoise}$ for time $\tau$ between two pulses as
\begin{align}
    &\mathcal T \exp[-\mi \int_{t-\tau}^t dt \qty(\hat H_{\rm mol} + \hat H_{\rm MWnoise}(t))]
    \nonumber
    \\
    =&
    \exp[-\mi \hat S^{(z)} \int_{t-\tau}^t dt h(t)]\exp(-\mi \tau \hat H_{\rm mol})
    \, .
\end{align}

Instead, of assuming Markovian white noise, which leads to a decay rate that is linear in time, we find that a simple Gaussian model with a time-independent variance captures the observed dynamics. More explicitly, if we define, $\int_{t-\tau}^t dt' h(t') \equiv \Delta\phi r_t$, with  $r_t$  a random variable, we can capture the observed dynamics. Because such  a phase kick occurs after the free evolution, we can absorb it into the pulse for our computations.

\subsection{Fitting the Model Parameters}
\label{ssec:theory_fitting}
\begin{table}[]
    \centering
    \begin{tabular}{cccc}
        States affected     & Parameter           & Value      & Section \\
        \hline
        Rotational States & $J$ & $h \times 36$ Hz & \ref{ssec:theory_tweezer} \\
        &$\delta/(2\pi)$            & $-500 \dots 500$ Hz     & \ref{ssec:theory_tweezer}  \\
        & $\gamma_{\rm deph}$ & 2.4 s$^{-1}$ & \ref{ssec:theory_microwave} \\
        & $\Delta\phi$        & $8 \times 10^{-3}$  & \ref{ssec:theory_microwave} \\
        \hline
        Hyperfine States & $\gamma_{\rm hf}$   & 2 s$^{-1}$ & \ref{ssec:theory_noninteracting} \\
        & $\Delta\Omega \,\tau$      & $2 \times 10^{-3}$ & \ref{ssec:theory_noninteracting}
    \end{tabular}
    \caption{Decoherence model parameters.}
    \label{tab:decoherence}
\end{table}

\begin{figure*}
    \centering
    \includegraphics[width=\linewidth]{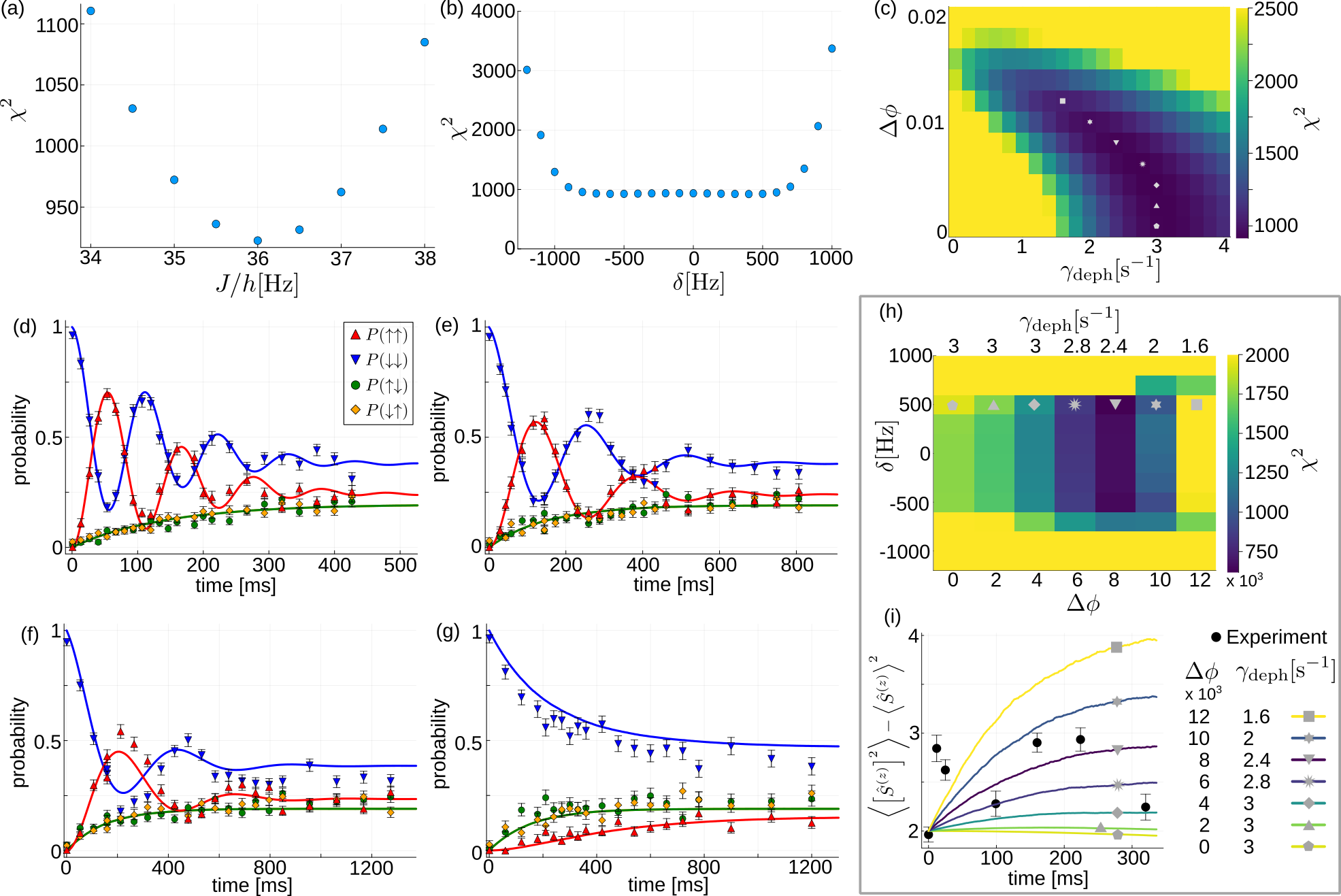}
    \caption{Determining Model Parameters from Two-Molecule Data.
    (a) Sum of residuals $\chi^2$ for simulations using different values of $J$. The optimum appears at $J/h=36$\,Hz. Other parameters are $J/h=36$\,Hz, $\delta = 2\pi \times 400$\,Hz, $\gamma_{\rm deph} = 2$\,s$^{-1}$ and $\Delta\phi = 10^{-2}$.
    (b) Sum of residuals $\chi^2$ for simulations using different detunings $\delta$. A wide region of good fits is observed, originating from the robustness of XY8 and Floquet XXZ sequences to static detunings.
    (c) Sum of residuals for a joint fit of $\Delta\phi$ and $\gamma_{\rm deph}$. The color axis is truncated above $\chi^2 = 2500$.
    (d,e, f,g) Fits to the two-body interaction data at $J=36$ Hz, $\delta = 2\pi \times 400$ Hz, $\gamma_{\rm deph} = 2$\,s$^{-1}$ and $\Delta\phi = 10^{-2}$. The anisotropies are $\Delta = $ (d) 0.33, (e) 0.66, (f) 0.8, (g) 1.  
    (h) $\chi^2$ of the variance $\left\langle \qty[\hat S^{(z)}]^2 \right\rangle - \left\langle \hat S^{(z)} \right\rangle^2$ for one to 11 molecules spaced at 8.4\,$\mu$m, such that $J \approx 0$. The $x$-axis jointly tunes $\Delta\phi$ and $\gamma_{\rm deph}$ along the path of minimal $\chi^2$ as identified by the symbols in panel (c). The color axis is truncated above $\chi^2 = 2000$.
    (i) Example comparison of non-interacting spin variance between theory and experiment for $\mathcal{N}=8$ and $\delta = 2\pi \times 500$\,Hz. The symbol legend is identical to that of panel (h).
    }
    \label{fig:fitting}
\end{figure*}

To determine the parameters of our model, we measure the dynamics of two interacting molecules and compare them with theoretical predictions for different model parameters. Specifically, we vary the model parameters, and determine the optimal parameters by seeking the best fit between experiment and theory. We follow an iterative procedure, where we first fit $J$ with our initial guess of $\delta$, $\Delta\phi$, and $\gamma_{\rm deph}$, then fit $\delta$ with the updated $J = d^2/(4\pi\epsilon_0 a^3)$, and the initial $\Delta\phi$ and $\gamma_{\rm deph}$, and then jointly fit $\Delta\phi$ and $\gamma_{\rm deph}$ with $J$ and $\delta$.
To determine $\Delta\phi$ and $\gamma_{\rm deph}$, we additionally fit data taken from non-interacting molecules. We then repeat this process until all parameters are converged, i.e.~none of the guessed optimal parameters change after updating the other parameters, which happens after just a few iterations.
Throughout, we keep the motional temperature fixed to the experimentally determined value.

Schematically, we summarize the protocol as follows:
\begin{enumerate}
    \item Guess the parameters $J$, $\delta$, $\Delta\phi$, and $\gamma_{\rm deph}$.
    \item Optimize $J$ using the two-particle interacting data [Fig.~\ref{fig:fitting}(a)].
    \item Find a preliminary optimum for $\delta$ using the two-particle interactions [Fig.~\ref{fig:fitting}(b)].
    \item Find a cross-correlated region of confidence between $\gamma_{\rm deph}$ and $\Delta\phi$ [Fig.~\ref{fig:fitting}(c)].
    \item Run a 2D scan of the variance of non-interacting molecules for a joint scan of $\delta$ and $\Delta\phi$, using the optimal $\gamma_{\rm deph}$ determined in step 4 for each $\Delta\phi$. This fully constrains $\gamma_{\rm deph}$ and $\Delta\phi$ together with the previous step, and provides a more constrained range for $\delta$ [Fig.~\ref{fig:fitting}(h)].
    \item Restart from step 2. with the new optimal $J$, $\delta$, $\Delta\phi$, and $\gamma_{\rm deph}$.
\end{enumerate}

The fitting procedure is illustrated for the final parameters in Fig.~\ref{fig:fitting}. For $J$, we find a clear optimum around $J/h = 36$\,Hz. For $\delta$, we initially find a wide region of confidence $-900\,\text{Hz} < \delta < 700\,\text{Hz}$, which is wider than the $\sim500$\,Hz variations we expect experimentally.
For $\gamma_{\rm deph}$ and $\Delta\phi$, we find a correlated region of good confidence [Fig.~\ref{fig:fitting}(c)]. To constrain $\delta$, $\Delta\phi$, and $\gamma_{\rm deph}$ further, we measure the evolution of the variance $\left\langle \qty[\hat S^{(z)}]^2 \right\rangle - \left\langle \hat S^{(z)} \right\rangle^2$ of non-interacting molecules for different molecule numbers $1 \leq N \leq 11$ and compare them to numerical results for different $\delta$ while varying $\Delta\phi$ and $\gamma_{\rm deph}$ along the minimum. Crucially, the collective variance is more sensitive to global noise $\Delta\phi$ than to noise acting individually on each molecule $\gamma_{\rm deph}$, which allows us to determine $\Delta\phi$ precisely.
The results are shown in Fig.~\ref{fig:fitting}(h).
We find a clear optimum for $\Delta\phi \approx 8 \times 10^{-3}$ and $\gamma_{\rm deph} \approx 2.4$\,s$^{-1}$, with $\abs{\delta}/(2\pi) \lesssim 500$\,Hz. This detuning range is consistent with day-to-day drift and calibration errors of the microwave resonance. Furthermore, Fig.~\ref{fig:fitting}(i) shows example traces at $\delta/(2\pi) = 500$\,Hz, compared to the experimental data.
The simulations clearly capture the general trend of growing variance for growing $\Delta\phi$, and the microwave phase noise described by $\Delta\phi$ clearly improves the fit quality. The optimal parameters are summarized in Tab.~\ref{tab:decoherence}.

\subsection{Modeling Decoherence and Loss in the Non-interacting Storage Manifold}
\label{ssec:theory_noninteracting}
During storage in the non-interacting hyperfine states, the dynamical decoupling is achieved by optical 2-photon Raman pulses instead of a direct microwave transition.
Consequently, we use a slightly different error model to account for different physical error sources. Instead of dephasing, we find that fully depolarizing noise, modeled by three different single-molecule jump operators, $L_{i,\alpha} = \sqrt{\gamma_{\rm hf}} \hat s^{(\alpha)}_i/2$, better captures the single-particle noise.  The model reproduces experimental  measurements with $T_{1,\text{hf}} \approx T_{2,\text{hf}}$. The rate of this noise remains unchanged at $\gamma_{\rm hf} = 2$\,s$^{-1}$.
Furthermore, we observe that additional phase kicks are not needed to explain decoherence. We therefore set $\Delta\phi_{\rm hf} = 0$. We instead include amplitude noise, a known imperfection, which is modeled by
\begin{align}
    \hat{\mathcal U}_{\rm Raman}(r_t') &=
    \exp[-\mi \Omega \tau \qty(1 + \Delta\Omega r_t') \hat S^{(\alpha)}]
    \, .
\end{align}
Here, $r_t'$ is a Gaussian random variable with $\overline{r_t'} = 0$ and $\overline{r_t' r_{t'}'} = \delta_{tt'}$, and $\tau\Delta\Omega = 2\times 10^{-3}$ is the noise amplitude. We attribute these fluctuations to the fact that the Raman Rabi frequency $\Omega$ depends on the laser detuning as well as the laser Rabi frequencies. The level of fluctuations in Rabi frequency is consistent with an independent probe of the shot-to-shot fluctuation in the Raman pulse power. 

\subsection{Numerical Details}
\label{ssec:numerical_details}

\begin{figure}
    \centering
    \includegraphics[width=\columnwidth]{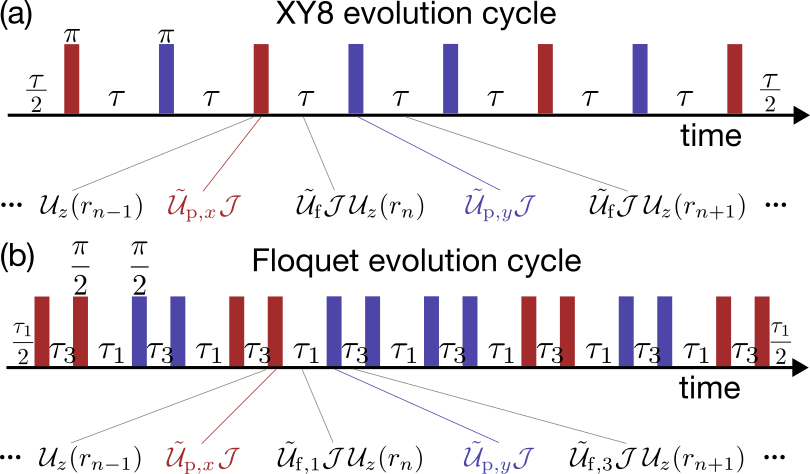}
    \caption{XY8 and Floquet Pulse Sequences. (a) XY8 sequence. (b) XY8-XXZ Floquet sequence. For simulations, each pulse and each step between the pulses are described by non-unitary evolution operators given by Eqs.~\eqref{eq:theory_pulse} to \eqref{eq:theory_f2}. Here, $\tilde{\mathcal U}_{\mathrm f}$ in the XY8 sequence is analogous to $\tilde{\mathcal U}_{{\mathrm f},i}$ with $\tau_i = \tau$. Each non-unitary evolution is followed by a jump $\mathcal J$. Finally, before each pulse, we apply a random $z$-rotation $\hat{\mathcal U}_z(r_n)$.}
    \label{fig:theory_sequences}
\end{figure}

We simulate the dynamics in Julia~\cite{bezanson2017julia}. To simulate time evolution of the Master equation, we use quantum trajectories~\cite{daley2014quantum} for $\mathcal{N}=8$ molecules and full master equation evolution for the two-molecule simulations. We combine the trajectory sampling with sampling the thermal initial state and sampling the microwave phase kicks with size $\Delta \phi \, r_t$.

We now explain the simulations and provide an example of time evolution of the Floquet dynamics. The evolution during XY8 dynamical decoupling is fully analogous, both for molecules in two different rotational levels and in the same rotational level but in  two different   hyperfine manifolds. First, we draw a set of motional harmonic oscillator quantum numbers,  $n^{(i)}_\alpha$, and compute the Hamiltonian Eq.~\eqref{eq:h_mol} with:  site disorder given by Eq.~\eqref{eq:delta_i}, nearest-neighbor interactions given by Eq.~\eqref{eq:Jij_neighbor} and long-range interactions given by Eq.~\eqref{eq:Jij_perturbative}.
We then compute the non-unitary evolution operators,
\begin{align}
    \tilde{\mathcal U}_{{\rm p}, \alpha} &=
    \exp[-\mi \tau_{\rm p} \qty(\hat H_{\rm mol} + \hat H_{\rm MW, \alpha} - \frac{\mi}{2} \sum_k \hat L_k^\dagger \hat L_k)]
    \, ,
    \label{eq:theory_pulse}
    \\
    \tilde{{\mathcal U}}_{\rm f,1} &=
    \exp[\frac{-\mi \tau_1}{2} \qty(\hat H_{\rm mol} - \frac{\mi}{2} \sum_k \hat L_k^\dagger \hat L_k)]
    \, ,
    \label{eq:theory_f1}
    \\
    \tilde{\mathcal U}_{\rm f,2} &=
    \exp[-\mi \tau_2 \qty(\hat H_{\rm mol} - \frac{\mi}{2} \sum_k \hat L_k^\dagger \hat L_k)]
    \, .
    \label{eq:theory_f2}
\end{align}
During the squeezing dynamics, we only include jump operators, $\hat L_k$, associated with dephasing, while for the hyperfine dynamics, we include both dephasing and blackbody radiation. Furthermore, we pre-compute $\hat{\mathcal U}_z(r)$ for $-3 \leq r \leq 3$ in steps of 0.2. For hyperfine noise, we similarly pre-compute $\hat{\mathcal U}_{\rm Raman,\alpha}(r)$ for $-3 \leq r \leq 3$ in steps of 0.2. Note that for the Floquet sequence, $\tilde{\mathcal U}_{\rm p,\alpha}$ describes a $\pi/2$ pulse, but for the XY8 sequence, $\tilde{\mathcal U}_{\rm p, \alpha}$ is a full $\pi$ pulse.

Then, we iteratively evolve the state $\ket \psi$ according to the following loop: 
\begin{enumerate}
    \item Compute $\ket{\psi'} = \tilde{\mathcal U}_\xi \ket\psi$ for the appropriate $\xi$.
    \item Apply the jump gate $\mathcal J$ as follows: \begin{enumerate}
        \item Draw a uniformly distributed random number $0 \leq p \leq 1$.
        \item Check whether $\bra{\psi'}\ket{\psi'} < p$?
        \item If ``Yes,'' randomly choose the appropriate quantum jump and apply it (see below)~\cite{daley2014quantum};\\
        If ``No,'' renormalize the state $\ket{\psi'}$
    \end{enumerate}
    \item Repeat.
\end{enumerate}

To choose the appropriate jump, we compute the jump weights $w_k = \bra{\psi'}\hat L_k^\dagger \hat L_k \ket{\psi'}$, such that the corresponding probabilities are $p_k = w_k / (\sum_k w_k)$.
Then, to apply the jump, we compute $\ket{\psi''} = \hat L_k \ket{\psi'}/\sqrt{w_k}$.
The implementation of this algorithm is illustrated in Fig.~\ref{fig:theory_sequences}.

To increase efficiency, we reuse the same thermal distribution and thus the same $\hat{\mathcal U}$ for multiple runs (10 for Floquet squeezing dynamics, 40 for XX squeezing dynamics) before moving to the next thermal ensemble. In total, we use 10,000 trajectories, which is sufficient for convergence. For the two-particle simulations, we follow essentially the same procedure, but we do not use quantum trajectories, because the additional sampling costs exceed the cost increase due to simulating a full density matrix for two molecules instead of two state vectors. In that case, we vectorize the density matrix and replace $\tilde{\mathcal U}_{\rm f/(p,\alpha)} \rightarrow \exp[\mathcal L_{\rm f/(p,\alpha)}\tau_{\rm f/(p,\alpha)}]$, and apply the microwave phase noise as in Eq.~\eqref{eq:theory_uz_densitymatrix}.
This absorbs the jump channel $\mathcal J$ in the evolution.

When simulating the blackbody radiation dynamics, we only simulate coherent dynamics in the $\ket\uparrow$-$\ket\downarrow$ manifold. All incoherent dynamics following a BBR event is simulated in post-processing.

\subsection{Modeling Imperfections of Rotational to Hyperfine Transduction}
In this section, we give details on how imperfections during transduction from rotational encoding to hyperfine encoding are modeled.

\subsubsection{Correlated Global Rotations}
\label{sec:gaussian_noise_model}
To model imperfections during transduction, we consider a globally correlated random rotation of the collective spin $\mathbf{S}=\langle {\hat {\mathbf{S}}}\rangle$. The noise is modeled as a shot-to-shot random unitary rotation on the collective spin operators,
\begin{equation}
	\hat{U}_{\mathrm{noise}} = e^{-i \boldsymbol{\theta}\cdot \hat{\mathbf{S}}},
\end{equation}
where the rotation angles are drawn from a Gaussian distribution with zero mean,
\begin{equation}
	\overline{ \theta_i } = 0,
	\qquad
		\overline{ \theta_i \theta_j } = \sigma^2 \delta_{ij}.
\end{equation}
Here, $\sigma$ parametrizes the width of the correlated rotation noise.

For sufficiently small rotation angles, the collective spin operators transform as
\begin{equation}
	\mathbf{S}' \approx \mathbf{S} + \boldsymbol{\theta}\times \mathbf{S}.
\end{equation}

We consider an initial state polarized predominantly along the $y$-axis, such that $\langle \hat S_y \rangle \gg \langle \hat S_x \rangle,\langle \hat S_z \rangle$. Under isotropic Gaussian noise, the mean spin length decays according to
\begin{equation}
	\langle \hat S_y \rangle
	=
	\langle\hat  S_y \rangle_0 e^{-\sigma^2},
\end{equation}
where the factor of two in the exponent relative to single-axis noise arises because fluctuations about both the $x$ and $z$ axes depolarize the collective spin.

The variance along the squeezed quadrature is similarly modified by the noise process. Assuming independent and zero mean Gaussian noise, we find
\begin{equation}
	\mathrm{Var}(\hat S_z')
	\approx
	\mathrm{Var}(\hat S_z)_0
	+
	\sigma^2 \langle \hat S_y^2 \rangle
	+
	\sigma^2 \langle \hat S_x^2 \rangle.
\end{equation}

For a state strongly polarized along $y$, the dominant contribution arises from the coherent spin projection,
\begin{equation}
	\langle \hat S_y^2 \rangle
	\approx
	\langle \hat S_y \rangle^2,
\end{equation}
such that
\begin{equation}
	\mathrm{Var}(\hat S_z')
	\approx
	\mathrm{Var}(\hat S_z)_0
	+
	\sigma^2 \langle \hat S_y \rangle^2.
\end{equation}

\subsubsection{Noise-Modified Theory Curves and Fit Procedure}
\begin{figure}
	\centering
 	\includegraphics[width=\columnwidth]{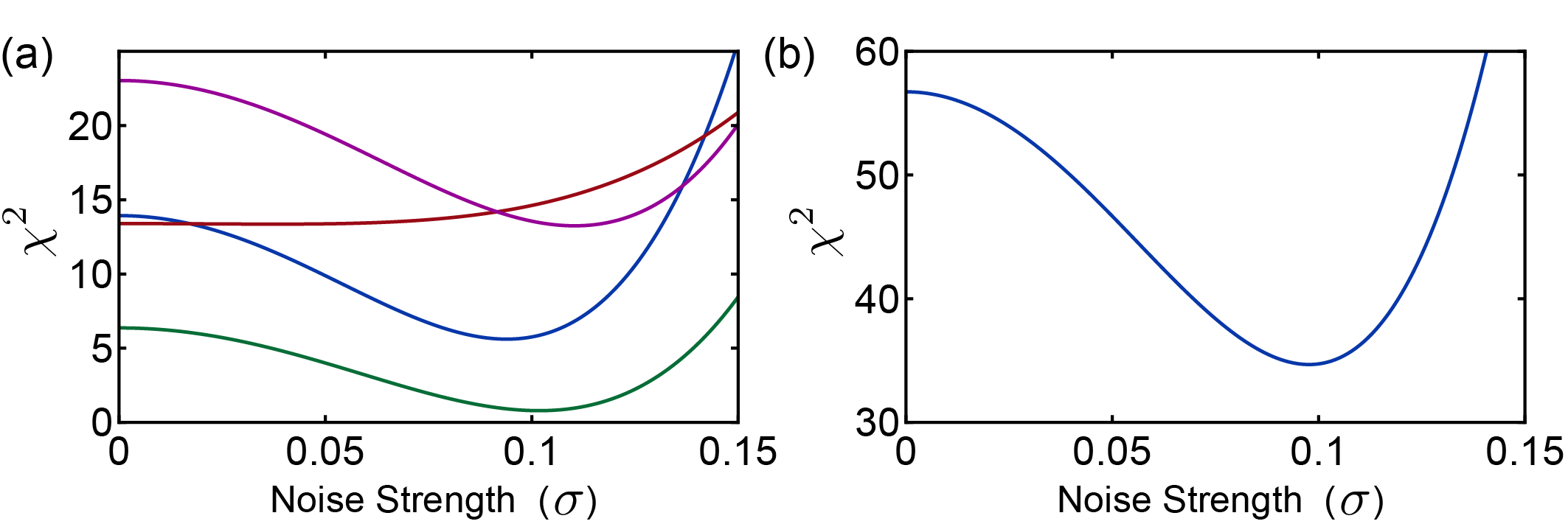}
	\caption{Determining Strength of Transduction Noise. (a) Shown are the fit $\chi^2$ versus the Gaussian noise width $\sigma$ for the squeezed state spin variance (blue), squeezed state spin length (red), coherent spin state spin variance (green), coherent spin state spin length (purple). All show mutually compatible optima at $\sigma\approx0.1$ (b) The combined $\chi^2$ versus Gaussian noise width $\sigma$. This is used to determine the strength of the Gaussian noise $\sigma_\text{opt}$.}
	\label{fig:transductionchisq}
\end{figure}

When comparing between theory and experiment, we account for correlated rotation noise during transduction by modifying theory curves computed according to the equations above. Specifically, we multiply the collective spin lengths by the Gaussian suppression factor, so that
\begin{equation}
	\langle \hat S_y \rangle_{\mathrm{th}}
	\rightarrow
	\langle \hat  S_y \rangle_{\mathrm{th}} e^{-\sigma^2}.
\end{equation}
The variances are modified according to
\begin{equation}
	\Delta \hat S_z^2{}_{\mathrm{th}}
	\rightarrow
	\Delta \hat S_z^2{}_{\mathrm{th}}
	+
	\sigma^2  \langle \hat S_y \rangle_{\mathrm{th}}^2.
\end{equation}
The Gaussian noise width $\sigma$ is treated as a free fit parameter and extracted by simultaneously fitting four experimental observables:
\begin{enumerate}
	\item The squeezed-state spin length,
	\item The squeezed-state spin variance,
	\item The coherent-spin-state (CSS) spin length,
	\item The coherent-spin-state (CSS) variance.
\end{enumerate}
For each data set, we perform a self-consistency check by independently determining the best-fit Gaussian noise width $\sigma$ using a reduced chi-squared cost function,
\begin{equation}
	\chi^2(\sigma)
	=
	\sum_i
	\frac{
		\left[
		y_i^{(\mathrm{exp})}
		-
		y_i^{(\mathrm{th})}(\sigma)
		\right]^2
	}{
		\delta y_i^2
	},
\end{equation}
where $y_i^{(\mathrm{exp})}$ are the measured data points, $y_i^{(\mathrm{th})}(\sigma)$ are the corresponding noise-modified theory predictions, and $\delta y_i$ are the experimental uncertainties. As shown in Fig.~\ref{fig:transductionchisq}(a), the four independent fits give mutually compatible best fit noise strengths.

To determine the best fit Gaussian noise strength, the combined total cost function is computed as
\begin{equation}
	\chi^2_{\mathrm{tot}}
	=
	\chi^2_{\mathrm{sq,var}}
	+
	\chi^2_{\mathrm{sq,len}}
	+
	\chi^2_{\mathrm{css,var}}
	+
	\chi^2_{\mathrm{css,len}}.
\end{equation}
Fig.~\ref{fig:transductionchisq}(b) shows the combined $\chi^2$ from which the optimal noise strength is determined. We find $\sigma_\text{opt}=0.098$, corresponding to $\sim5^\circ$ global noise rotations. This is consistent with independent estimates of rotational angle errors due to amplitude noise in our pulses.

\section{Theory Predictions for Scalable Squeezing in 2D}
While spin squeezing from dipolar exchange interactions in 1D is generally limited to a few dB even for long chains, in 2D, we expect the spin squeezing to scale with the number of molecules~\cite{Perlin2020spinsqueeze}. However, whether this survives realistic decoherence and disorder remains to be investigated.

Here, we apply our model for decoherence and disorder to 2D geometries. Because our model has been validated in this work with experimental data in a 1D setting, we believe that it should accurately predict performance in 2D settings.

\begin{figure}
	\centering
	\includegraphics[width=0.8\linewidth]{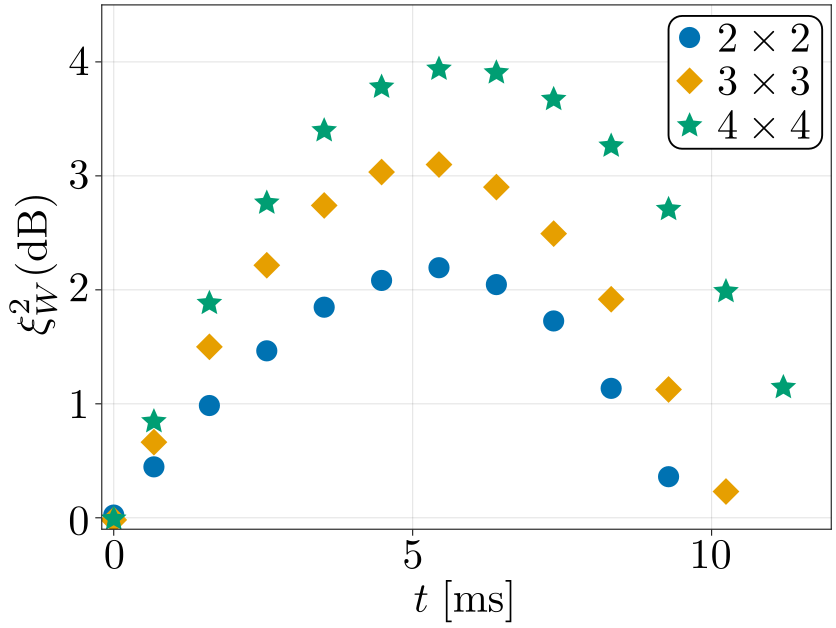}
	\caption{Scalable Spin-squeezing in 2D Square Lattices. Numerical simulations for different lattice sizes using the noise model presented here. The parameters used are: lattice filling $f=0.9$, $\Delta\phi = 8\times 10^{-3}$, $\gamma_{\rm deph} = 2.4 \,\text{s}^{-1}$.
	}
	\label{fig:2D}
\end{figure}

Specifically, we investigate whether scalable squeezing is observable using realistic decoherence and disorder parameters identified in our 1D investigations.
We simulate the squeezing dynamics of up to 16 molecules in a square geometry with our noise model. Informed by experimental capabilities, we assume a filling fraction of $f = 0.9$. Simulation results, shown in Fig.~\ref{fig:2D}, show clear scaling of spin squeezing with system size. We anticipate that with ongoing investigations on preparing larger arrays, these system sizes (and beyond) will be accessible in the near future.

In the following sections, we discuss the different noise sources and how they reduce the achievable spin squeezing in these 2D square geometries.

\subsection{Interaction Disorder}
In contrast to a unit-filled square lattice ($f=1$), sub-unit filling $f < 1$ introduces disorder in the interaction strength. As discussed by Kwasigroch and Cooper in Refs.~\cite{kwasigroch2014bose,kwasigroch2017synchronization}, the molecular rotations remain in phase in the presence of interaction disorder due to sub-unit filling $f<1$, as long as $f > f_c \approx 0.15$. For $f < f_c$, the array decomposes into many small clusters with strong internal interactions and weak coupling to the environment. With filling fractions of $f = 0.9$ (demonstrated already in our experiment), we are deeply within this spin-ordered phase, where scalable spin squeezing is expected to persist.

In addition to the lattice filling, the finite motional temperature adds further disorder to the interaction strength as discussed in Sec.~\ref{ssec:theory_tweezer}.
Compared to the previous section, the equation describing $J_{ij}$ is modified because the quantization axis has to be out-of-plane to create $1/r^3$ interactions. Following the same approach used to derive Eq.~\eqref{eq:Jij_perturbative}, for a quantization axis along $z$, we find
\begin{widetext}
\begin{align}
    J_{ij}\qty(\vec n^{(i)}, \vec n^{(j)})
    =
    \frac{d^2}{4\pi\epsilon_0 R^3}
    &\left[
    1
    +
    \frac{\qty(15\frac{R_x^2}{R^2}-3)a_{\mathrm{ho},x}^2}{R^2} 
    \qty(n_x^{(i)} + n_x^{(j)} + 1)
    \right.
    \nonumber
    \\
    &\quad
    \left.
    +
    \frac{\qty(15\frac{R_y^2}{R^2}-3)a_{\mathrm{ho},y}^2}{R^2} 
    \qty(n_y^{(i)} + n_y^{(j)} + 1)
    -
    \frac{9a_{\mathrm{ho},z}^2}{R^2}
    \qty(n_z^{(i)} + n_z^{(j)} + 1)
    \right]
    \, .
    \label{eq:Jij_2D}
\end{align}
\end{widetext}
Here, we set $\vec R = (R_x, R_y, 0)^T$ with $\abs{\vec R} = R$ is the distance vector connecting sites $i$ and $j$. For numerical simulations, we restrict $n_z \leq 50$ to remain in the regime of validity of this approximation. For thermal values $\bar n_z \approx 26$, we do not expect this restriction or the quadratic approximation to qualitatively affect any results. For typical thermal values, the corrections to $J_{ij}$ are dominated by the $z$-component, reaching around
\begin{align}
    \frac{9a_{\mathrm{ho},z}^2}{a^2} (2\bar n_z + 1)
    \approx
    0.62
    \, ,
\end{align}
for nearest neighbors. While this value indicates that we are at the edge of the perturbative regime, perturbation theory consistently underestimates the interaction strength, thus providing an effectively lower estimate of the achievable spin squeezing. The reduction in interaction strength is comparable to the interaction-strength difference $1 - \frac{1}{2^{3/2}} \approx 0.65$ between nearest and next-nearest neighbors in 2D. Crucially, even if we assume that all molecules with $\bar n_z > 26$ are effectively removed from the system, this would only remove about $1/e \approx 37\%$ of the molecules, still leaving us well within the spin-aligned regime for scalable spin squeezing. Our numerical simulations shown in Fig.~\ref{fig:2D} confirm this.

We note that this type of finite temperature interaction disorder can be reduced by cooling the molecules further, for example, via Raman sideband cooling~\cite{caldwell2020sideband,Lu2024RSC,Bao2024RSC}.

\subsection{Decoherence}
Since the disorder on the interaction strength is sufficiently small such that the spins remain approximately aligned, we can estimate the effect of the different decoherence channels in the collective spin manifold. Here, we give a succinct treatment of noise in the collective manifold. A more detailed analysis can be found in the Supplemental Material of Ref.~\cite{chu2021quantum}.

In the collective manifold, the system is described by a one-axis twisting Hamiltonian
\begin{align}
    \hat H = \chi \qty[\hat S^{(z)}]^2
    \, ,
\end{align}
where $\chi = \frac{1}{\mathcal{N}} \sum_{j \neq i} J_{ij} \sim \frac{4f J}{\mathcal{N}}$ is the average interaction strength, which depends on the actual molecule number $\mathcal{N}$, the precise molecule distribution and thermal occupation numbers.
This Hamiltonian achieves optimal spin-squeezing at $t_{\rm opt} = \frac{\hbar}{\chi} 3^{1/6} \mathcal{N}^{-2/3} \sim 0.3\mathcal{N}^{1/3} \hbar / (f \times J) \approx N^{1/3} \times 1.6$\,ms\cite{ma2011quantum} for $J = h \times 33$\,Hz, where this value takes into account thermal averaging which reduces the interaction strength from its bare value of $h\times 36\,\text{Hz}$.

Single-particle dephasing leads to a decay in the spin vector and an increase in variance, which reduces the precision gain by a factor of $\xi^2 / \xi^2_{\rm opt} = 1 + 3\gamma_{\rm deph} t_{\rm opt} \approx 1 + 0.012 \mathcal{N}^{1/3}$.
While this does affect the large-$\mathcal{N}$ scaling behavior, even up to $\mathcal{N} = 100$, this reduces squeezing by only about -0.2\,dB at the optimal squeezing time of about 7\,ms.

In addition to single-particle dephasing, the microwave phase noise described by Eq.~\eqref{eq:theory_uz_definition} also reduces the contrast and increases the variance.
This can be described in the Heisenberg picture as
\begin{align}
    &\int dr \frac{\exp(-r^2/2)}{\sqrt{2\pi}} \hat{\mathcal U}_z(r)^\dagger \hat S^{(x)} \hat{\mathcal U}_z(r)
    = \exp(-\frac{\Delta\phi^2}{2})\hat S^{(x)}
    \, ,
    \\
    &\int dr \frac{\exp(-r^2/2)}{\sqrt{2\pi}} \hat{\mathcal U}_z(r)^\dagger \qty[\hat S^{(y)}]^2 \hat{\mathcal U}_z(r)
    \nonumber
    \\
    &\quad=
    \frac{\cosh(\Delta\phi^2) \qty[\hat S^{(y)}]^2 + \sinh(\Delta\phi^2) \qty[\hat S^{(x)}]^2}{\exp(\Delta\phi^2)}
    \, ,
    \\
    &\int dr \frac{\exp(-r^2/2)}{\sqrt{2\pi}} \hat{\mathcal U}_z(r)^\dagger \qty[\hat S^{(y)} \hat S^{(z)}] \hat{\mathcal U}_z(r)
    \nonumber
    \\
    &\quad=
    \exp(-\frac{\Delta\phi^2}{2})\hat S^{(y)}\hat S^{(z)}
    \, .
\end{align}
Interestingly, with respect to squeezing, phase kicks applied at a rate $1/\tau$ act like global noise with Lindblad operator $\sqrt{\Gamma}\hat S^{(z)}$ with $\Gamma = \Delta\phi^2/\tau \approx 0.16$ s$^{-1}$ for small $\Delta\phi \ll 1$. Following Ref.~\cite{chu2021quantum}, this increases the squeezing by a factor $\xi^2 / \xi^2_{\rm opt} = 1 + 2N\Gamma t$. Again, this type of behavior does change the large-$N$ scaling behavior, but for $N=100$, and keeping the experimental sequence with pulse spacings of about $400\,\mu\text{s}$, the squeezing is only decreased by $\approx-0.9$ dB, on top of 10\,dB of squeezing with 100 particles for $1/r^3$ interactions in 2D~\cite{Perlin2020spinsqueeze}.

\clearpage

\section{Supplementary Data}

\begin{figure*}
	\centering
	\includegraphics[width=\linewidth]{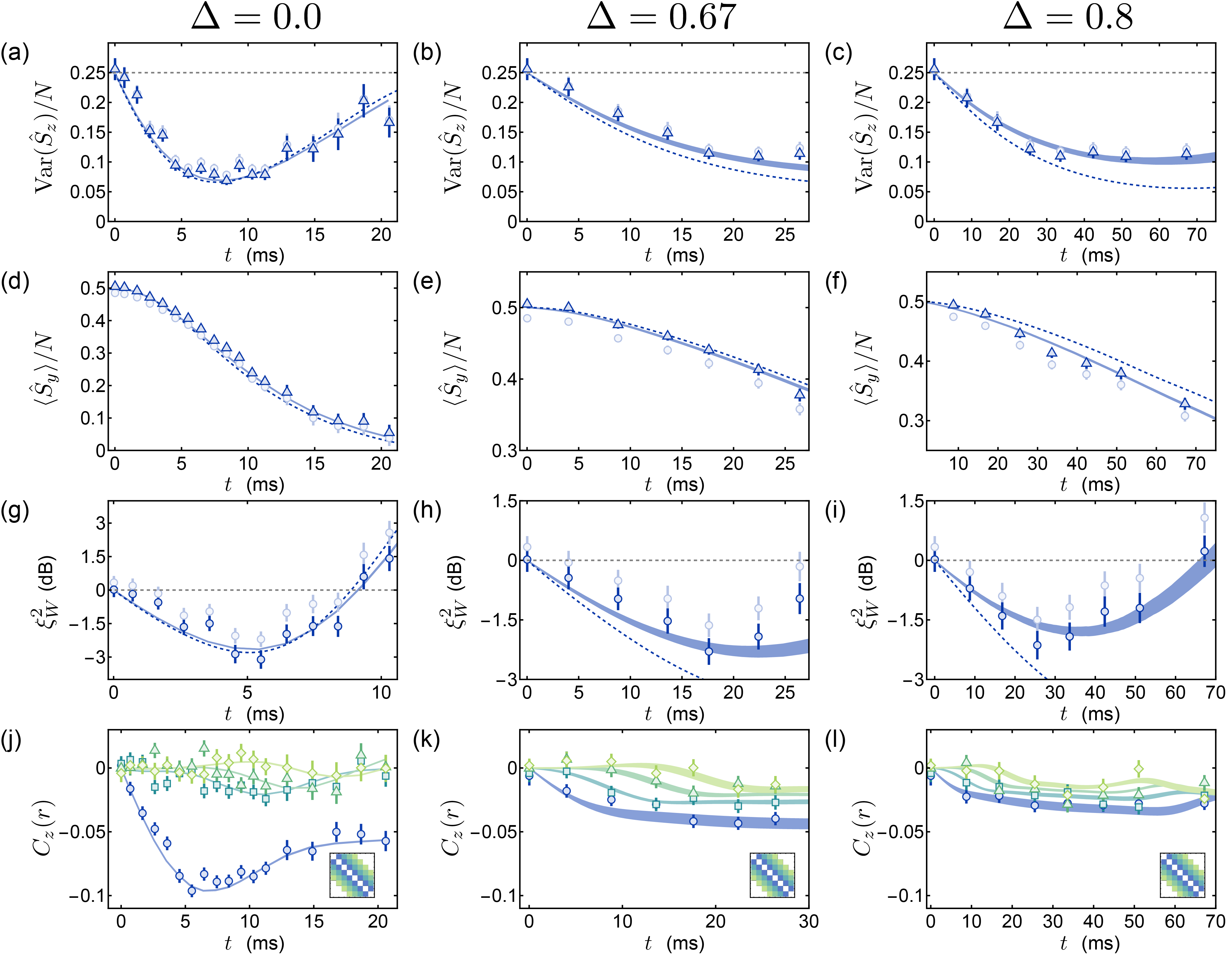}
	\caption{Spin Squeezing Dynamics. (a,b,c) Normalized spin variance $\text{Var}[\hat{S}_z]/\mathcal{N}$ (a,b,c) versus squeezing duration $t$ for the Ising strengths $\Delta=0.0$, $0.67$, and $0.80$, respectively. (d,e,f) Normalized spin length $|\hat{S}_y|/\mathcal{N}$ versus squeezing duration $t$ for the Ising strengths $\Delta=0.0$, $0.67$, and $0.80$, respectively.  (g,h,i) Wineland parameters $\xi^2_W$ versus squeezing duration $t$ for the Ising strengths $\Delta=0.0$, $0.67$, and $0.80$, respectively. (j,k,l) Mean spatial spin-spin correlator $C_z(r)$ versus squeezing duration $t$ for the Ising strengths $\Delta=0.0$, $0.67$, and $0.80$, respectively. $r$ denotes the spin separation in number of sites. For all plots, solid (light) data points are with (without) measurement correction. Ideal XXZ theory is shown by dashed lines. Our full model including decoherence and disorder is shown by the colored bands.}
	\label{fig:alldata_sqdynamics}
\end{figure*}

\begin{figure*}
	\centering
	\includegraphics[width=\linewidth]{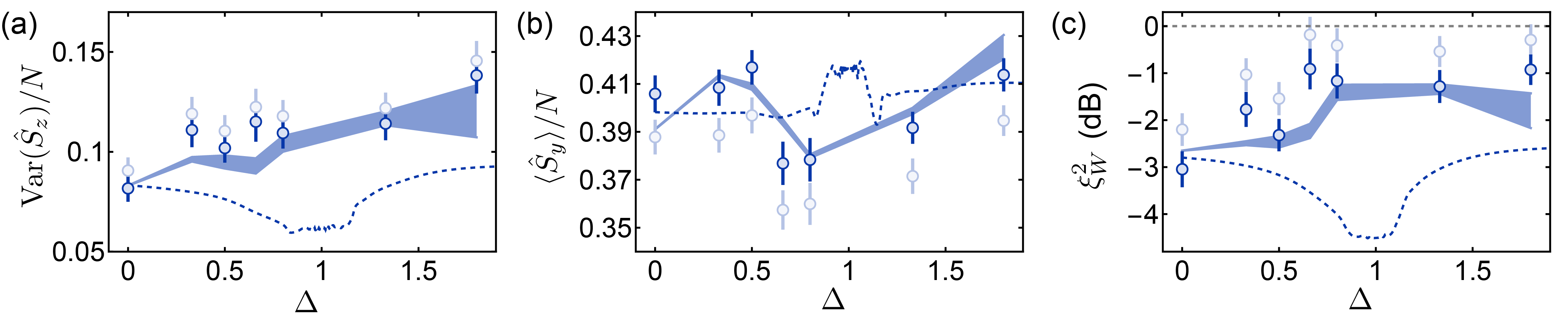}
	\caption{Squeezed State Properties versus Ising Strength ($\Delta$). (a) Normalized spin variance $\text{Var}[\hat{S}_z]/\mathcal{N}$ versus Ising anisotropy $\Delta$. (b) Normalized spin length $|\hat{S}_y|/\mathcal{N}$ versus Ising anisotropy $\Delta$. (c) Wineland parameter $\xi^2_W(t^*)$ versus Ising anisotropy $\Delta$. For all plots, solid (light) data points are with (without) measurement correction. Ideal XXZ theory is shown by the dashed lines. Our full model including decoherence and disorder are shown by the colored bands.}
	\label{fig:alldata_anisotropy}
\end{figure*}

\begin{figure*}
	\centering
	\includegraphics[width=\linewidth]{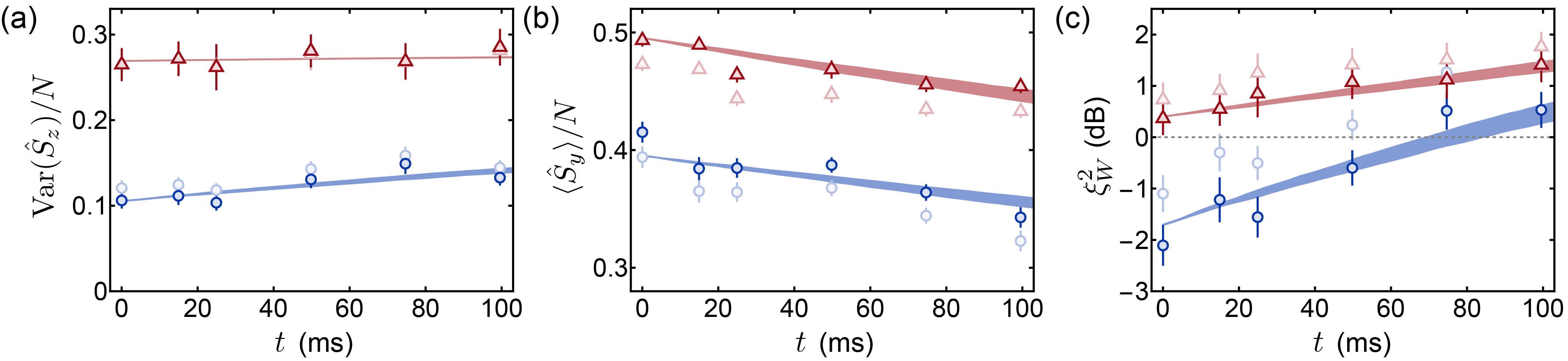}
	\caption{Properties of Hyperfine-Transduced States vs Storage Duration. (a) Normalized spin variance $\text{Var}[\hat{S}_z]/\mathcal{N}$ versus storage duration $t$. (b) Normalized spin length  $|\hat{S}_y|/\mathcal{N}$ versus storage duration $t$. (c) Wineland parameter $\xi^2_W(t^*)$ versus storage duration $t$. For all plots, data for the XX squeezed state and the coherent spin state are shown by the blue circles and red triangles, respectively. Solid (light) data points are with (without) measurement correction. The colored bands show our full theory model with decoherence, disorder and transduction errors.}
	\label{fig:alldata_storage}
\end{figure*}

In this section, we provide supplementary data for the spin-squeezed states investigated in the main text. Fig.~\ref{fig:alldata_sqdynamics} shows squeezing dynamics at three Ising strengths $\Delta=0.0$, $0.67$, and $0.80$. Specifically, the quantities of normalized spin variance, spin length, Wineland parameter, and the connected spin-spin correlator $C_z(r)$ versus interaction time are provided, in analogy to Fig.~2(b,c,d) and Fig.~3(d,e) in the main text. Fig.~\ref{fig:alldata_anisotropy} shows the normalized spin variance, spin length, and Wineland parameter of squeezed states (also shown in the main text and obtained from the first two quantities) as a function of Ising strength $\Delta$. These are for spin-squeezed states created with the no-noise optimal squeezing duration $t*$. Fig.~\ref{fig:alldata_storage} shows the normalized spin variance, spin length and Wineland parameter for hyperfine storage as a function of storage duration. These complement data shown in Fig.~4(e) of the main text. Properties for both the transduced spin-squeezed state and the coherent spin state are shown. These plots also show predictions from ideal $1/r^3$ XXZ model and also our theory that includes decoherence and disorder. We observe good agreement between experiment and our full model.

%